\documentclass[%
 reprint,
superscriptaddress,
 amsmath,amssymb,
 aps,
]{revtex4-2}

\usepackage{braket}

\usepackage{graphicx}% Include figure files
\usepackage{dcolumn}% Align table columns on decimal point
\usepackage{bm}% bold math
\usepackage[dvipsnames]{xcolor}

\usepackage[export]{adjustbox}
\usepackage{floatrow}
\usepackage[caption=false]{subfig}
\usepackage{comment}

\usepackage{wasysym} % for the diameter command 

\begin{document}

% \addbibresource{biblio.bib}

% \preprint{APS/123-QED}

\title{Kilometer-Scale Ion-Photon Entanglement with a Metastable $^{88}$Sr$^{+}$ Qubit}

% Direct Infrared Ion-Photon Entanglement On the Kilometer Scale Using a Metastable $^{88}$Sr$^{+}$ Qubit

\author{Mika A. Zalewski}
    % \email{Second.Author@institution.edu}
    \affiliation{%
 Duke Quantum Center, Departments of Electrical and Computer Engineering and Physics, Duke University, Durham, NC 27708
}%
\author{Denton Wu}
\affiliation{%
 Duke Quantum Center, Departments of Electrical and Computer Engineering and Physics, Duke University, Durham, NC 27708
}%
\author{Ana Luiza Ferrari}
\affiliation{%
 Duke Quantum Center, Departments of Electrical and Computer Engineering and Physics, Duke University, Durham, NC 27708
}%
\author{Yuanheng Xie}
\affiliation{%
 Duke Quantum Center, Departments of Electrical and Computer Engineering and Physics, Duke University, Durham, NC 27708
}%
\author{Norbert M. Linke}
\affiliation{%
 Duke Quantum Center, Departments of Electrical and Computer Engineering and Physics, Duke University, Durham, NC 27708
}%
\affiliation{Joint Quantum Institute, University of Maryland, College Park, MD 20742, USA}
\email{norbert.linke@duke.edu}

\date{\today}

\begin{abstract}
We demonstrate entanglement between the polarization of an infrared photon and a metastable $^{88}$Sr$^+$ ion qubit. This entanglement persists after transmitting the photon over a $2.8\:$km long commercial fiber deployed in an urban environment. Tomography of the ion-photon entangled state yields a fidelity of $0.949(4)$  within the laboratory and $0.929(5)$ after fiber transmission, not corrected for readout errors. Our results establish the strontium ion as a promising candidate for metropolitan-scale quantum networking based on an atomic transition at $1092\:$nm, a wavelength compatible with existing telecom fiber infrastructure. 
\end{abstract}

\maketitle

\section{\label{sec:intro}Introduction}

Quantum networks consist of physically separated local quantum systems, or nodes, entangled via photonic interconnects \cite{qinternet, interconnect_overview}. 
They allow for distributed quantum computing, forming a path to scalability by connecting separate quantum processors into a larger computing system \cite{Monroe_interconnects, neutral_interconnects_overview, Covey2023_review}. 
Additionally, quantum networks with links beyond the laboratory scale have applications in quantum sensing \cite{telescope_repeaters, Nichol2022}, blind quantum computing \cite{Drmota_blindQC, Fitzsimons2017_blindQC_overview}, quantum cryptography \cite{crypto, diqkd_theory, Nadlinger2022, Zhang2022_diqkd}, and fundamental tests of quantum mechanics \cite{Hensen2015, bell_test_atoms}. 

Trapped ions \cite{inns2024_101km, Kucera2024_14km_network}, neutral atoms \cite{Ritter2012, vanLeent2022_neutrals_33km}, color centers \cite{stolk_dist_classical_network, Knaut2024_bostonnetwork}, and quantum dots \cite{quantumdots} are common choices for quantum network nodes. Trapped ions are also among the leading qubit platforms due to their long coherence times and high fidelity gates \cite{longest_coherence, smith2025_single_qubit_uwave, löschnauer2024_best2qubit_ion}. Using trapped ions, photonic interconnects have been demonstrated with high entanglement fidelities as well as high rates of entanglement generation \cite{Main2025, Saha2025, Oreilly_symp_cool}.

 Optical fibers are commonly used to transmit photons in a quantum network link. The majority of available atomic transitions are resonant with light in the visible to ultraviolet spectrum, frequencies with high transmission losses in silica fibers. 
 In trapped ions, work with infrared transitions at $854\:$nm can enable longer, direct networks \cite{inns_remote_ent}, though such wavelengths are still lossy at kilometer-scale distances.
 One method to improve transmission is quantum frequency conversion (QFC), which transfers the photon state to a frequency compatible with existing telecom fibers \cite{Ikuta2011, Geus24_hansonQFC, Bock2018, Quraishi2023_QFC}. Demonstrations of long-distance remote entanglement have been reported using QFC, but conversion losses limit the process \cite{stolk_dist_classical_network, Kucera2024_14km_network, inns2024_101km}. 
%However, the conversion process decreases entangled state fidelities and rates 
These losses are caused by inefficiencies of both device and waveguide coupling as well as noise from background pump photons, which result in current conversion efficiencies of less than $0.60$ \cite{QFC_eff_17, QFC_25, QFC_eff_46, QFC_eff_57, Kucera2024_14km_network}.
% state-of-the-art end-to-end conversion efficiencies are limited to the $17-57\%$ range \cite{QFC_eff_17, QFC_25, QFC_eff_46, QFC_eff_57}. The wide range of values showcases the difficulty of an efficient implementation, which would be necessary at every node of the network, due .  

% Recently, initial demonstrations of generating entanglement in the infrared directly 
Recently, initial demonstrations of generating entanglement in the telecom-band directly
have been shown with neutral atoms and solid state platforms within a laboratory setting \cite{covey_telecom_network, erbium, direct_spin}. In this paper, we demonstrate a novel direct infrared entanglement scheme using a metastable $^{88}$Sr$^{+}$ qubit, and the transmission of the photons through a field-deployed fiber. The $1092\:$nm transition from the $5P_{1/2}$ to the $4D_{3/2}$ level in strontium exhibits a loss of $0.7\:$dB/km in SMF-$28$ fiber compared to the $0.2\:$dB/km at the optimal telecom wavelength of $1550\:$nm \cite{telecom}.
Given the tradeoff between this higher loss and the losses of QFC, transmission at $1092\:$nm is favorable at intermediate ($<20\:$km) node distances.

The emission of $1092$ nm photons leaves the  $^{88}$Sr$^{+}$ ion in a superposition of two metastable qubit (m-qubit) states. We discuss state preparation, measurement, and coherent operations for this new qubit. We demonstrate high-fidelity generation of an entangled state between this m-qubit and the polarization qubit of the photon both in the laboratory and over a $2.8\:$km field-deployed fiber. 

\section{ION-PHOTON ENTANGLEMENT}

The experiment is conducted using a $^{88}$Sr$^+$ ion confined in a linear Paul trap consisting of four rods and two hollow end caps. We employ this design due to its large trap depth and high optical access. The secular frequencies are $(\omega_x,\omega_y,\omega_z)=2\pi\times(1.23, 1.36, 0.15)\:$MHz. 
 The ion is optically pumped at $422$ and $1092\:$nm to initialize in $|5S_{1/2},+1/2\rangle$, and then excited to $|5P_{1/2}, -1/2\rangle$ using an $18\:$ns pulse at $422\:$nm. 
 Decay along five possible channels generates an entangled state between all available photon states and the corresponding electronic states, with a $5.6\%$ chance of decaying along the $1092\:$nm transition \cite{UDportal}. 

\floatsetup[figure]{style=plain,subcapbesideposition=top}
\begin{figure}    
    \sidesubfloat[]{%
        \includegraphics[width=0.433\linewidth, valign=m]{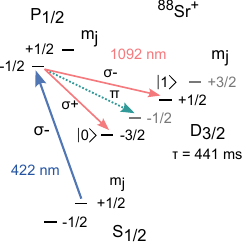}%
        \vphantom{ \includegraphics[width=0.42\linewidth,valign=m]{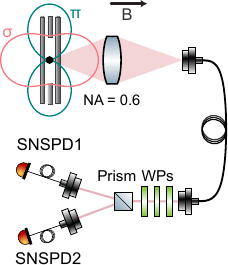}}
    }
    \hspace{-2mm}
    \sidesubfloat[]{%
        \includegraphics[width=0.43\linewidth,valign=m]{Figures/simple_layout_vert.pdf}%
    }
    \caption[j]{Quantum networking scheme. a) Reduced level diagram in $^{88}$Sr$^+$. The atomic qubit states $|0\rangle$ and $|1\rangle$ are $|4D_{3/2},-3/2\rangle$ and $|4D_{3/2},+1/2\rangle$ respectively. A short pulse of $\sigma^-$ light at $422\:$nm excites population from $|5S_{1/2},+1/2\rangle$ to $|5P_{1/2},-1/2\rangle$. Ion-photon entanglement at $1092\:$nm is generated upon decay to $4D_{3/2}$. b) Experimental setup. The emission patterns of the $\pi$ and $\sigma$ polarizations are shown with the ion at the center. By symmetry, $\pi$ light is not collected into the fiber, since the collection optics are parallel to the magnetic field. The photons are sent through SMF-$28$ optical fiber of length $2\:$m or $2.8\:$km.
    A series of three waveplates ($\lambda/4$, $\lambda/2$, and $\lambda/4$) provides full control over the polarization state of the photon. A Wollaston prism acts as a beam splitter before two detectors (SNSPDs). 
    }
    \label{fig:setup}
\end{figure}
 
 A custom objective with a numerical aperture (NA) of 0.6 collects only the $1092\:$nm photons along the quantization axis given by an external magnetic field and couples them into SMF-28 fiber. The emission patterns of $\pi$ versus $\sigma^+$ and $\sigma^-$ light allow us to collect only $\sigma$ light into the fiber with this geometry \cite{luo2009protocols}. We measure the lower bound of peak fiber coupling efficiency to be 66\%, achievable due to the relaxed alignment tolerances required for coupling long-wavelength photons.
 A quarter wave-plate maps $\sigma^+$ and $\sigma^-$ light to the states $|H\rangle$ and $|V\rangle$, respectively, before measurement. The resulting entangled state is given by
\begin{equation}
\label{eqn:en-state}
    |\psi\rangle=\frac{\sqrt{3}}{2}|H\rangle|0\rangle+\frac{1}{2}|V\rangle|1\rangle,
\end{equation}
where the atomic qubit states $|0\rangle$ and $|1\rangle$ correspond to the Zeeman states $|4D_{3/2}, -3/2\rangle$ and $|4D_{3/2}, +1/2\rangle$ respectively, as shown in Fig. \ref{fig:setup}a. 
% The photonic qubit states $|H\rangle$ and $|V\rangle$ correspond to the two polarization directions, which are generated using a quarter wave-plate before detection. to the two circularly polarized  
The imbalanced amplitudes in Eqn. \ref{eqn:en-state} result from the unequal Clebsch-Gordon coefficients of the two decay channels. 
While this state is not maximally entangled, it can be used to generate a Bell-state between two distributed ions, at the expense of a decrease in the entanglement generation rate \cite{type_2_theory, Stute2012}.  

\begin{figure}
    \centering
    \includegraphics[width=1\linewidth]{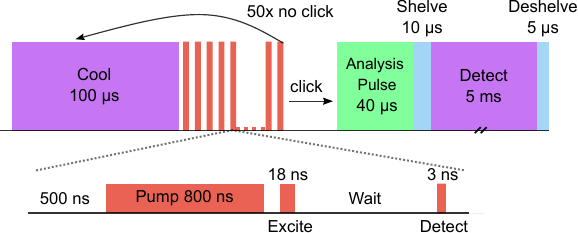}
    \caption{Experimental sequence for generation of ion-photon entanglement. After laser cooling, up to $50$ attempts of entanglement generation are made. Each includes optical pumping to $|5S_{1/2},+1/2\rangle$ and then excitation to $|5P_{1/2},-1/2\rangle$. The wait time before detection depends on the length of fiber. If a photon click is registered on one of the two detectors during the detection window, the sequence proceeds to ion state rotation and measurement. Each attempt includes about $500\:$ns of latency. }
    \label{fig:exp}
\end{figure}

The full experimental sequence to generate and measure ion-photon entanglement is shown in Fig. \ref{fig:exp}. This sequence contains a loop for fast, consecutive attempts at entanglement generation, which breaks either upon detection of a photon for measuring the ion state, or every $50$ cycles for ion cooling.
The collected photons are sent via optical fiber to a polarization analysis setup, shown in Fig. \ref{fig:setup}b, which includes a polarizing beam splitter and two superconducting nano-wire single photon detectors (SNSPDs).
A series of motorized waveplates before the beam splitter is used to measure the photon along different bases. Entanglement is heralded upon detection of a photon on one of the detectors. 

Because the excitation pulse is driven by an acousto-optic modulator (AOM), the pulse length is longer than the lifetime of the $P_{1/2}$ level. 
In this work, we limit the photon detection window to mitigate any effect from re-excitation.
If performing remote ion-ion entanglement, a faster excitation pulse, such as from a pulsed laser, would be necessary to avoid a decrease in the fidelity.

\section{\label{sec:exp}METASTABLE ION QUBIT}

The ion qubit states can be rotated using a pair of Raman beams detuned by $17.3\:$GHz from the  $4D_{3/2}$ to $5P_{3/2}$ transition at $1004\:$nm. 
The fidelity of a Raman $2\pi$-pulse is measured to be $0.9865(9)$, after correcting for readout error, which is limited by off-resonant scattering. This fidelity could be improved by increasing the detuning and laser power.

The ion state is measured via electron shelving.
Optical pumping transfers the population from $|1\rangle$ to the $4D_{5/2}$ level such that subsequent fluorescence detection at $422\:$nm results in a bright ion if the state was in $|0\rangle$ and a dark ion if it was in $|1\rangle$. In our readout scheme, leakage errors out of the qubit manifold also appear as dark events during fluorescence detection, since all population in the $5S_{1/2}$ and $4D_{3/2}$, except $|4D_{3/2},-3/2\rangle$, is shelved. Results are post-selected on instances in the qubit manifold by performing each experiment in two separate sets of trials. In the first set of trials, the bright state is taken to be $|0\rangle$. In the second set of trials, a Raman $\pi$-pulse is performed before detection in order to map the $|1\rangle$ state to bright. We keep only instances of bright state detection in both cases; states outside of the qubit manifold will remain dark even after the $\pi$-pulse. Further details of this readout scheme are discussed in Appendices A and B. 

For future experiments, a narrow linewidth laser could be used for single-shot readout.
This laser would allow us to directly shelve both qubit states to $4D_{5/2}$ to check for leakage errors prior to qubit state measurement. In both readout schemes, qubit leakage errors result in a decrease in entanglement generation rate but not fidelity.

\section{Results}

\subsection{Entanglement Generation Rate}

We utilize two optical fibers for different demonstrations of ion-photon entanglement. The first is a $2\:$ m SMF-$28$ Ultra fiber for tests within our laboratory. The second is a $2.8\:$km SMF-28e$+$ optical fiber loop that is deployed in the field, running underground in downtown Durham, NC. 
The rate of entanglement generation is set by both the attempt rate and success probability.
The total duration of the entanglement generation loop is $2.136\:\mu$s with the $2\:$m fiber. For long-distance fiber networks, the attempt rate is dominated by travel time in the fiber, in our case, an additional $13.613\:\mu$s \cite{inns2024_101km, Ruskuc2025_multiplexing, duan_multiplex}.
This corresponds to an attempt rate of $468,165$/s in the laboratory, and an attempt rate of $63,496$/s with the $2.8 \:$km fiber.

Experimentally, we measure an entanglement generation success probability of $7.64(8)\times10^{-4}$ with a $20\:$ns detection window through the short fiber, resulting in an initial entanglement generation rate of $350(4)$/s. 
The random photon arrival time within the detection window causes phase uncertainty in the ion qubit. Therefore, when measuring the ion in the X- or Y-basis, we use a shorter detection window of $3\:$ns, resulting in a rate of $96(1)$/s. 
Attenuation in the long fiber decreases our success probability to $2.57(6)\times10^{-4}$ with a $15\:$ns detection window.  Combined with the considerably slower attempt rate as 
a result of the photon travel time, this gives an entanglement generation rate of $15.9(4)$/s. 

The total probability of successful entanglement generation is given by
\begin{equation}
\label{rate}
    P_{ent} = P_p \cdot P_c \cdot P_q \cdot P_w,
\end{equation}
where $P_p$ is the likelihood of emitting the desired photon and $P_c$ is the likelihood of the photon reaching the detector, including all fiber coupling and transmission losses.
$P_q$ is the quantum efficiency of the detector, and $P_w$ accounts for the finite detection window. In our experiment, $P_p$ is $0.056$, given by the branching ratio of the $1092\:$nm transition \cite{UDportal}. $P_cP_q$ is $0.0168(3)$. $P_w$ is $>0.85$ for a $20\:$ns detection window and is $0.18$ for a $3\:$ns detection window, showing that there is a rate tradeoff when improving the fidelity by shortening the detection window.

The exact success probability can be measured directly from the entanglement generation attempts and successes. However, these successes will include instances where population has left the qubit manifold. Therefore, the success probabilities are adjusted to reflect only successful entanglement generation attempts, where the ion is in the qubit manifold. The likelihood of a trial being outside the qubit manifold is measured to be $0.0197$ for the laboratory data and $0.0234$ for the deployed fiber data, with a slight variation as data is taken on different days. 

We obtain $P_cP_q$ above with an experiment to directly measure the detection efficiency of the high NA imaging system. We turn on the $422$ nm $\sigma^+$ and $\sigma^-$ beams together, without a $1092\:$nm beam on, guaranteeing emission of a single photon at $1092\:$nm each experimental cycle. We then detect this photon and measure the total number of successful detections compared to the total number of experimental cycles to get the detection efficiency.

The $0.6$ NA objective only covers $10\%$ of solid angle. 
Additionally, in the current setup, the transmission through the analysis optics is $\sim0.65$, which can be overcome in future work through improved alignment or in-fiber optics. The lower bound on the fiber coupling, $0.66$, is measured directly after optimization of the fiber alignment. This efficiency decreases over the span of several hours and is optimized several times during each experimental run.
The SNSPD (Quantum Opus, Opus One) has a measured quantum efficiency of $\sim0.80$ at $1092$ nm.

\floatsetup[figure]{style=plain,subcapbesideposition=top}
\begin{figure}
    \sidesubfloat[]{%
    \includegraphics[width=0.9\linewidth, valign=m]{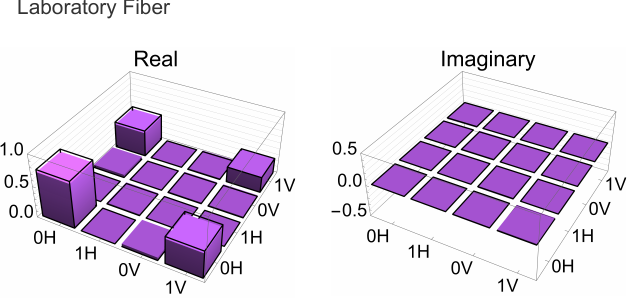}% 
    }\vspace{2mm}
    \sidesubfloat[]{%
        \includegraphics[width=0.9\linewidth,valign=m]{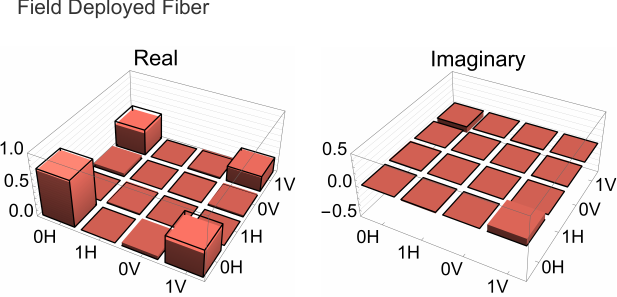}%
    }
    \caption{Results. The density matrix is reconstructed using Maximum Likelihood Estimation for the ion-photon entangled state through (a) $2\:$m of optical fiber in the laboratory, resulting in a fidelity of $0.949(4)$ and a purity of $0.908(7)$, and (b) $2.8\:$km of optical fiber deployed underground, resulting in a fidelity of $0.929(5)$ and a purity of $0.899(9)$. No active polarization stabilization is used for the deployed fiber. Due to the imbalanced coefficients in the entangled state, the diagonal elements are not equal. Ideal values are shown as outlines on the bars.}
    \label{fig:density_matrices}
\end{figure}

\subsection{Laboratory Fiber}

The fidelity of the experimentally generated entangled state with respect to the target state is determined using quantum state tomography. Measurements 
in the  three Pauli bases are performed on each qubit for a total of nine measurement settings, as detailed in Appendix C. The density matrix, $\rho$, can be reconstructed directly from these measurements, but the result is not guaranteed to be positive semi-definite, due to experimental errors. Therefore, we use a constrained Maximum Likelihood Estimation (MLE) to generate the final density matrix \cite{tomo_quant_inf, tomography}. Using $F=\langle\psi|\rho|\psi\rangle$ where $|\psi\rangle$ is given in Eqn. \ref{eqn:en-state}, we recover a fidelity of $0.949(4)$, with the full density matrix shown in Fig. \ref{fig:density_matrices}. The purity of the state, defined as $P = \text{Tr}(\rho^2)$, is $0.908(7)$. The purity sets an upper bound on the fidelity of $0.952(4)$. 
When the fidelity is at the bound, it is limited by decoherence mechanisms. 
However, if the fidelity is much lower, then this infidelity is due to an unwanted unitary \cite{purity}. In this case, the the result is limited by decoherence. Further details on this relation can be found in Appendix D.

\begin{figure}
    \centering
    \includegraphics[width=1\linewidth]{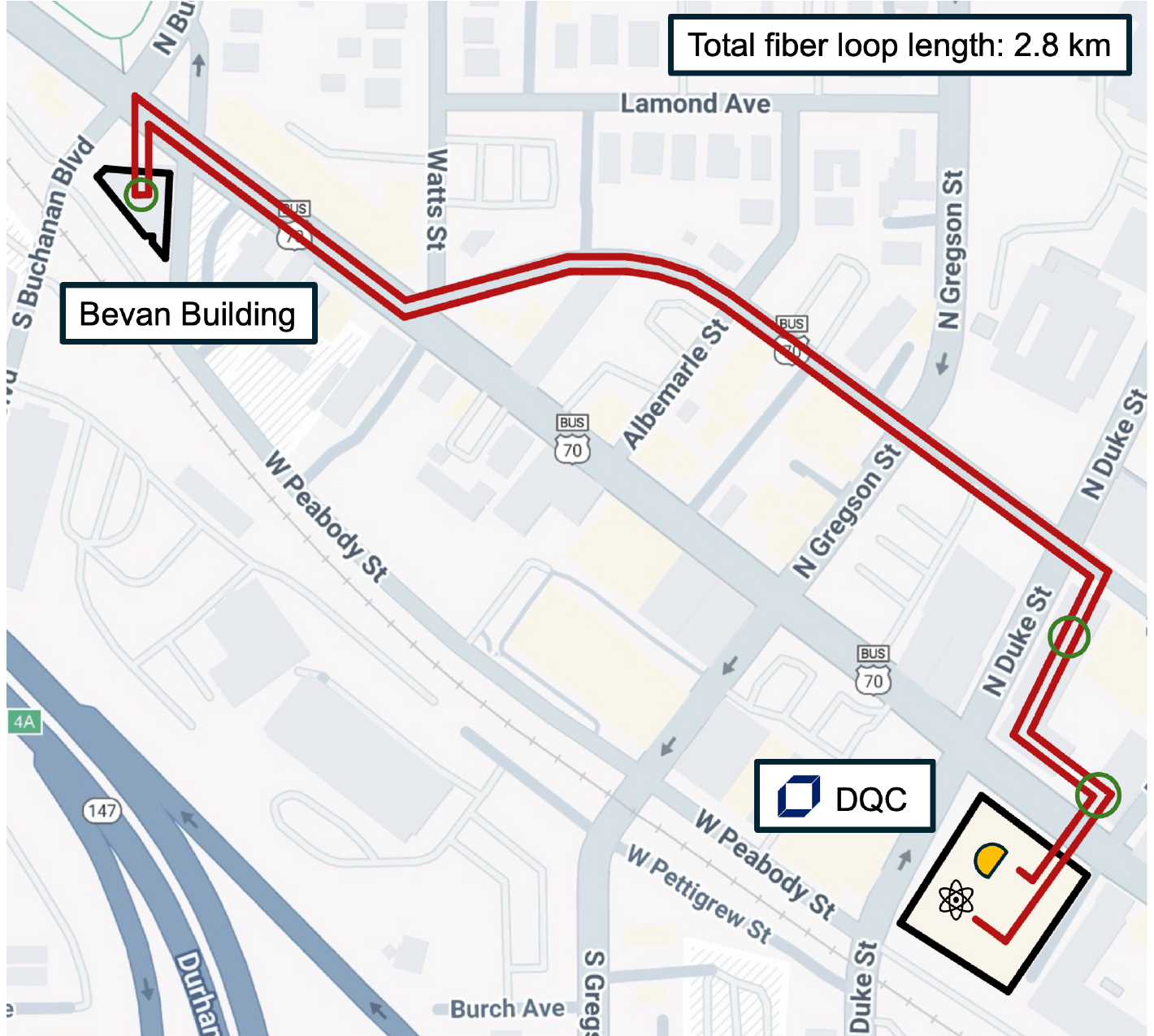}
    \caption{Map of the fiber loop in downtown Durham, NC. The laboratory is located at the Duke Quantum Center (DQC). The fiber runs underground except for the locations marked by the green circles, which are  network utility rooms. The fiber contains three splices and no connectors along its length.}
    \label{fig:FiberMap}
\end{figure}

Of the characterized errors, ion qubit decoherence due to magnetic field noise and ion qubit readout errors are the most significant. There are also a number of smaller contributions, such as those from errors in the photon path, predominantly due to imperfect waveplates, and errors in the $\pi/2$ pulses (see Table \ref{tab:error_table}). We attribute the remaining fidelity loss to polarization mixing in the imaging system \cite{Hucul2015}.
The polarization errors arise from inhomogeneous birefringence in the high NA vacuum window and imaging optics. 
The coherence time of the ion qubit is $1.36(6)\:$ms. This is limited by random magnetic field noise, leading to exponential decay in the coherences \cite{monz2011quantum}.
The largest readout error contribution is beam polarization error during electron shelving. We model the atomic structure using an Optical Bloch Equations simulation \cite{Hugothesis} to quantify how various experimental parameters affect the entangled state fidelity (see Appendix A). 
Additional sources of infidelity from detector background, imperfect unitary operations on both the ion and the photon qubit, and Raman phase noise are collectively estimated to be $<10^{-2}$.

\subsection{Field-Deployed Fiber Loop}

We demonstrate the viability of our scheme for metropolitan-scale quantum networking by sending the $1092\:$nm photons through a $2.8\:$km underground fiber running across downtown Durham, NC, shown in Fig. \ref{fig:FiberMap}. The fiber loops back into the lab to the same polarization state detector used in the short fiber case. 

The loop consists of Corning SMF-28e$+$ fiber, one of the most widely deployed in the world, demonstrating our scheme's compatibility with existing telecom infrastructure \cite{jung2009comparative}. Further characterization of this fiber can be found in Appendix E. Quantum state tomography is again used to characterize the entangled state. We achieve a fidelity of $0.929(5)$ and a purity of $0.899(9)$. Atomic qubit decoherence accumulated during the longer photon travel time accounts for an additional $0.0035$ loss in fidelity. Here, the measured purity sets an upper bound on the fidelity of $0.948(5)$.
Therefore, the larger decrease in fidelity compared to the decrease in purity results from an uncontrolled unitary rotation \cite{purity}. We attribute this to a drift in polarization rotation in the long fiber between calibration and data taking, primarily due to temperature variation throughout the day. The remaining decrease in fidelity, and corresponding decrease in purity, is attributed to a small amount of depolarization in the fiber due to fast polarization noise, likely caused by vibrations. 
The maximum time between initial calibration and completion of data collection is $6$ hours.
No active fiber stabilization is used, demonstrating the robustness of transmitting ion-photon entangled states underground in an urban environment.

% The fidelity decrease compared to the short fiber stems predominantly from polarization instability in the long fiber. This can be seen in the difference between the fidelity and purity here, where 
% Because the decrease in the purity is small, the upper bound on the fidelity is now $94.8\%$. The discrepancy between that bound and the measured fidelity results from an uncontrolled unitary rotation, which we attribute predominantly to an drift in the long fiber polarization between calibration an tomography \cite{purity}. 

% Additionally, the entangled state fidelity bounds were measured periodically over a 24-hour period to test how mechanical noise and temperature drift throughout the day affected the ion-photon state generation. We measure just the bounds and not full QST to minimize the time required for each data point (see Supplemental Material). 
% The fidelity bounds as a function of time are shown in figure SOMETHING. We find no visible drop (REVISIT THIS IF WE HAVE A MORE PRECISE GAUGE) in the fidelity over the 24-hour period. 

{\renewcommand{\arraystretch}{1.3}
\begin{table}[]

    \caption{Error contributions in the measured state fidelity for the laboratory fiber and additional errors for the deployed fiber.}
    \centering
\begin{tabular}{ |p{4cm}|p{4cm}|  }
 \hline
 \multicolumn{2}{|c|}{Error Budget, Laboratory Fiber} \\ 
 \hline
 Error Source&Fidelity Error\\
 \hline
 Polarization Mixing & $2-2.5\times10^{-2}$ \\
 Atomic Qubit Decoherence & $1.25(6)\times10^{-2}$ \\
 Atomic Qubit Readout & $6(2)\times10^{-3}$ \\
 Photon Path Errors & $<5\times10^{-3}$ \\ 
 $\pi/2$-pulse Errors & $1\times10^{-3}$ \\
 Background Counts & $<1\times10^{-3}$ \\
 Imaging Alignment & $<1\times10^{-3}$ \\
 Raman Phase Noise & $<1\times10^{-3}$ \\
 \hline
 \multicolumn{2}{|c|}{Additional Errors in Deployed Fiber} \\
 \hline
  Error Source& Fidelity Error\\
 \hline
 Polarization Instability  & $1.65\times10^{-2}$\\
 Atomic Qubit Decoherence & $3.5\times10^{-3}$ \\
 \hline
\end{tabular}

    \label{tab:error_table}
\end{table}}

\section{Outlook}

We demonstrate the viability of the $^{88}$Sr$^+$ m-qubit and associated $1092\:$nm transition for quantum networking. Additionally, we provide insights into the behavior of entangled states distributed across deployed fiber, which have relevance to a wide range of quantum networks. With the current polarization encoding of the photonic qubit, active polarization stabilization of the deployed fiber provides a solution to mitigate fidelity loss \cite{Kucera2024_14km_network, stolk_dist_classical_network}. Since polarization errors account for a large portion of our infidelity, alternative photon encodings can also be considered, such as time bin or frequency qubits \cite{Saha2025, Ruskuc2025_multiplexing, berkeley_freq_qubits, streed_freq_qubits}. 
For all networking systems, the attempt rate for long-distance entanglement is limited by the travel time of the photon in the fiber. This limitation can be mitigated by strategies such as temporal multiplexing, which allow for multiple attempts in quick succession, as demonstrated in similar setups \cite{inns2024_101km, Ruskuc2025_multiplexing, duan_multiplex}. 
While the success probability of entanglement generation in the current setup is fundamentally limited by the atomic branching ratio, the use of an optical cavity for Purcell enhancement of the desired transition can improve both this rate and the collection efficiency \cite{goodwin_cavity_proposal, inns2024_101km}. 
Overall, this work provides an important step towards a city-scale, infrared quantum network with strontium ions.
% \textcolor{red}{Indeed, the use of an optical cavity is likely critical for achieving sufficiently high photon collection in any trapped ion networking system.} 
% % With such improvement, the remaining limits to the success probability are technical in nature.
% \textcolor{red}{}
% Assuming cavity integration at the state-of-the-art level \cite{inns_good_cavity} for all three ion systems, \textcolor{red}{provides a simplifier metric for comparison between systems. While exact details may vary, it is, in principle, possible to achieve nearly equivalent levels of cavity enhancement in different platforms.}
% \textcolor{red}{Under this assumption,} our scheme is favorable compared with current QFC-based schemes in Ca$^+$ \cite{inns2023_repeater, Kucera2024_14km_network} and Ba$^+$ \cite{Quraishi2023_QFC} up to a node distance of $5$ and $19\:$km, respectively, making it a promising options for a metropolitan quantum network.
% With these solutions to improve the entanglement generation rate, our system, with its technical simplicity and lower cost compared to QFC, is a promising option for a metropolitan quantum network.

%\section{Acknowledgments}
\vspace{2.5mm}
\textit{Acknowledgments} - We thank Hugo Janacek for sharing his GLOBES DX code for the Optical Bloch equation simulations and for his advice and guidance. We also thank Thomas Kim, Raphael Metz, Ecem Nur Duman, Andrew Kille, and Marko Cetina for assistance with the experimental setup. We acknowledge support from the Duke Office of Information Technology, Tingjun Chen, and Zehao Wang on the deployed fiber setup.
We thank George Toh, Michael Straus, Joe \mbox{Britton}, and Jungsang Kim for helpful discussions, and George Toh additionally for reviewing the manuscript.
This work was supported by the Army Research Office (grants W911NF-19-10296, W911NF-17-S-0002-0, W911NF-19-20181, and W911NF-22-10032), the National Science Foundation Convergence Accelerator program (OIA-2134891), and the Software-Tailored 
Architecture for Quantum CoDesign (STAQ) Award (PHY-2325080), as well as the funding from Duke University under the Beyond-the-Horizon and DST-Launch initiatives.

%reset the fig counter and add an S to the SM figures
\renewcommand{\thefigure}{A\arabic{figure}}
\setcounter{figure}{0}

\section*{Appendix A: Readout Method and Fidelity} \label{sec:readout_appendix}

\begin{figure}
    \sidesubfloat[]{%
    \includegraphics[width=0.7\linewidth, valign=m]{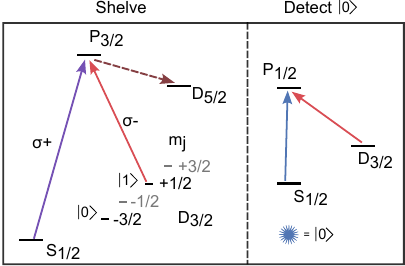}% 
    } \hfill \vspace{6mm}
    \sidesubfloat[]{%
        \includegraphics[width=0.9\linewidth,valign=m]{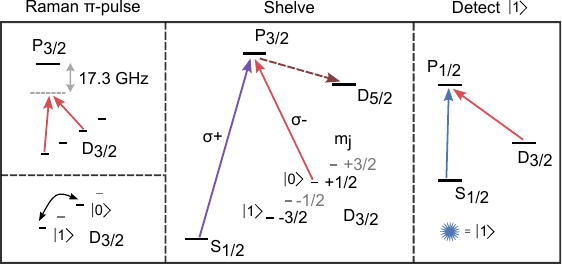}%
    }\hfill
    
    \caption{Qubit detection scheme. We use two different readout sequences for the two state populations. To detect the $|0\rangle$ state, shown in (a), we first optically pump the $|1\rangle$ state to $4D_{5/2}$, using a 408 nm $\sigma^+$ beam and a 1004 nm $\sigma^-$ beam. These beams also pump the $|4D_{3/2},-1/2\rangle$ and $|4D_{3/2},+3/2\rangle$ states to $4D_{5/2}$. We then turn on the 422 and 1092 nm lasers to detect the population that remains in $4D_{3/2}$, which is any population in $|0\rangle$. Detection of bright indicates $|0\rangle$ population. If the experimental errors that lead to population in $|4D_{3/2},-1/2\rangle$ and $|4D_{3/2},+3/2\rangle$ were negligible, detection of dark would indicate population in $|1\rangle$ with high fidelity. However, such leakage errors are non-negligible in our experiment. Thus, a separate detection method is required to measure $|1\rangle$. The experiment is re-run, this time with the measurement scheme shown in (b). A Raman $\pi$-pulse first flips the $|0\rangle$ and $|1\rangle$ states. The 408 and 1004 nm beams are then turned on to shelve. Finally, 422 and 1092 nm are used to detect the population remaining in $4D_{3/2}$, which this time corresponds to $|1\rangle$. In both sequences, population that has left the qubit manifold 
    is detected as dark. Therefore, by post-selecting on bright events only, we are able to exclude those errors.} 
    
    \label{fig:shelving}
\end{figure}

\begin{figure*}
    \centering
    \includegraphics[width=0.85\linewidth]{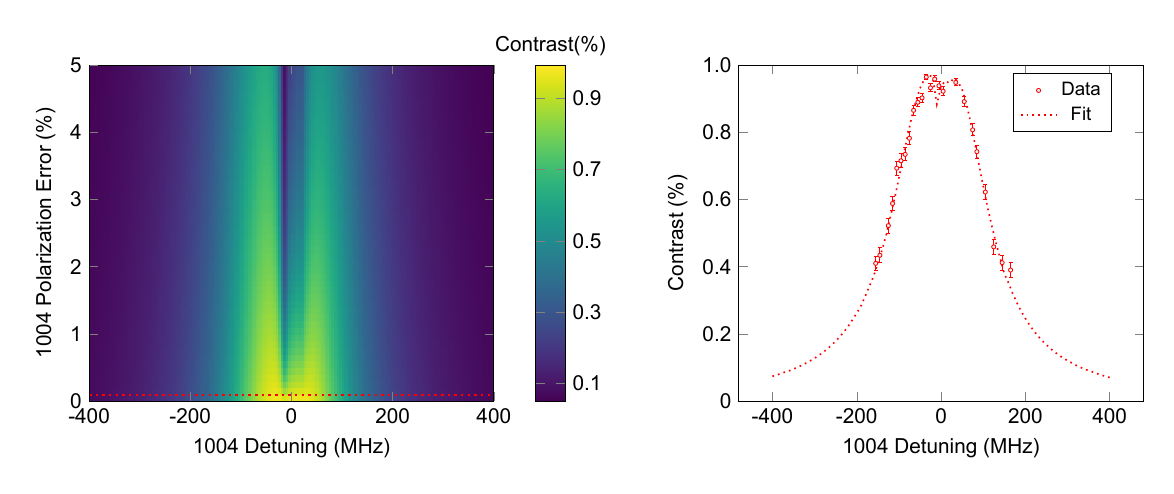}
    \caption{Simulation and experiment for determining $1004\:$nm polarization error. The shelving fidelity is a function of both 1004 detuning and polarization, as can be seen in the simulation on the left. The contrast is $F_{\text{bright}}+F_{\text{dark}}-1$ and is expected to be $1$ if all parameters are perfect. We fit an experimental scan of this contrast as a function of 1004 nm detuning, shown on the right. At each detuning, the optimal shelving time is recalibrated. Based on the experimental detuning fit, we can extract the $1004\:$nm polarization error from a cut of the simulation plot, which we find to be $9(1)\times10^{-4}$. This cut is shown as a dashed red line on the simulation plot.}
    \label{fig:shelving_characterization}
\end{figure*}

% With the readout errors fully characterized, we can incorporate them into the measurement matrix. 
Without error, we can express our readout scheme in the following manner. We perform two experiments, each with a total number of $k$ ion measurements, as shown in Fig. 5. In the first experiment, we shelve and then directly measure fluorescence, giving

\begin{equation} \label{eq:nopi_noerror}
    \begin{bmatrix}
        n_{b1} \\ n_{d1} \\ k 
    \end{bmatrix}
    =
    \begin{bmatrix}
        1&0&0\\0&1&1\\1&1&1
    \end{bmatrix}
    \begin{bmatrix}
        n_0\\n_1\\n_2
    \end{bmatrix},
\end{equation}
where $n_{b1}$ and $n_{d1}$ are the number of bright and dark occurrences, respectively, $n_0$ and $n_1$ are the number of occurrences in the $|0\rangle$ and $|1\rangle$ qubit states, respectively, and $n_2$ encompasses all cases where the ion state is outside of the qubit manifold. From this information alone, we cannot determine the separate numbers $n_0$, $n_1$, and $n_2$. Therefore, we repeat the experiment, using a $\pi$-pulse to swap the $|0\rangle$ and $|1\rangle$ states, giving

\begin{equation} \label{eq:pi_noerror}
    \begin{bmatrix}
        n_{b2} \\ n_{d2} \\ k 
    \end{bmatrix}
    =
    \begin{bmatrix}
        0&1&0\\1&0&1\\1&1&1
    \end{bmatrix}
    \begin{bmatrix}
        n_0\\n_1\\n_2
    \end{bmatrix},
\end{equation}
where $n_{b2}$ and $n_{d2}$ are the bright and dark occurrences for this second experiment. From these two experiments, we now have a set of five equations. With the constraints that $n_{b1}+n_{d1}=n_{b2}+n_{d2}=k$, we arrive at: 

\begin{equation}\label{readout}
    \begin{bmatrix}
        n_{b1} \\ n_{b2} \\ k 
    \end{bmatrix}
    =
    \begin{bmatrix}
        1&0&0\\0&1&0\\1&1&1
    \end{bmatrix}
    \begin{bmatrix}
        n_0\\n_1\\n_2
    \end{bmatrix}.
\end{equation}

Using this system of equations, we are able to analyze the bright counts from each experiment to find our total counts in each state. In order to understand the effect of shelving and readout errors on our qubit states, we next modify this set of equations to include the errors in the $\pi$-pulse and readout scheme described above. 

Equation \ref{eq:nopi_noerror} becomes 
\begin{equation} 
    \begin{bmatrix}
        n_{b1} \\ n_{d1} \\ k 
    \end{bmatrix}
    =
    \begin{bmatrix}
        1-\epsilon_b&\epsilon_d&\epsilon_{d2}
        \\\epsilon_b&1-\epsilon_d&1-\epsilon_{d2}
        \\1&1&1
    \end{bmatrix}
    \begin{bmatrix}
        n_0\\n_1\\n_2
    \end{bmatrix},
\end{equation}
where $\epsilon_b$ is the error from the qubit state $|0\rangle$ being measured as dark, $\epsilon_d$ is the error from $|1\rangle$ being measured as bright, and $\epsilon_{d2}$ is the error from population outside of the qubit manifold being measured as bright.

Equation \ref{eq:pi_noerror} is additionally modified to include the Raman $\pi$-pulse errors.  This pulse can have rotation errors, which we treat as bit flip errors. Additionally, it can have scattering errors that take population outside the qubit manifold. These errors result in the following matrix 
\begin{widetext}
\begin{equation}
\begin{split}
   & \begin{bmatrix}
        1-\epsilon_b&\epsilon_d&\epsilon_{d2}
        \\\epsilon_b&1-\epsilon_d&1-\epsilon_{d2}
        \\1&1&1
    \end{bmatrix}
    \begin{bmatrix}
        \epsilon_\pi&1-\epsilon_\pi&0
        \\1-\epsilon_\pi&\epsilon_\pi&0
        \\0&0&1
    \end{bmatrix}
    \begin{bmatrix}
        1-\epsilon_s&0&0
        \\0&1-\epsilon_s&0
        \\\epsilon_s&\epsilon_s&1
    \end{bmatrix} =
    \end{split}
\end{equation}

\begin{equation*}
    \begin{bmatrix}
         (\epsilon_d(1-\epsilon_{\pi})+\epsilon_{\pi}(1-\epsilon_b))(1-\epsilon_s)+\epsilon_{d2}\epsilon_s & ((1-\epsilon_b)(1-\epsilon_\pi)+\epsilon_d\epsilon_{\pi})(1-\epsilon_s)+\epsilon_{d2}\epsilon_s & \epsilon_{d2}
        \\((1-\epsilon_d)(1-\epsilon_{\pi})+\epsilon_{\pi}\epsilon_b)(1-\epsilon_s)+\epsilon_s(1-\epsilon_{d2})
        &(\epsilon_b(1-\epsilon_{\pi})+\epsilon_{\pi}(1-\epsilon_d))(1-\epsilon_s)+\epsilon_s(1-\epsilon_{d2})
        &1-\epsilon_{d2}
        \\1&1&1
    \end{bmatrix},
\end{equation*}
\end{widetext}
where $\epsilon_s$ is the scattering error from the Raman pulse, and $\epsilon_\pi$ is the rotation error in the Raman pulse. As $\epsilon_{d2}\epsilon_s<<1$, we neglect this term.
This gives the final version of Equation \ref{readout} with all errors as

\begin{widetext}
\begin{equation}
    \begin{bmatrix}
        n_{b1} \\ n_{b2} \\ k
    \end{bmatrix}
    = 
    \begin{bmatrix}
        1-\epsilon_b & \epsilon_d & \epsilon_{d2}\\
        (\epsilon_d(1-\epsilon_{\pi})+\epsilon_{\pi}(1-\epsilon_b))(1-\epsilon_s) & ((1-\epsilon_b)(1-\epsilon_\pi)+\epsilon_d\epsilon_{\pi})(1-\epsilon_s) & \epsilon_{d2} \\1&1&1
    \end{bmatrix}
    \begin{bmatrix}
        n_0 \\ n_1 \\ n_2
    \end{bmatrix}.
\end{equation}
\end{widetext}
Crucially, the errors in the experiment are sufficiently small that we can take $\epsilon_{d2}n_2<<1$, and we do not need to treat each state outside the qubit manifold individually. 
In our experiments, the errors are $\epsilon_b = 0.0159(5)$, $\epsilon_d = 0.005(2)$, and $\epsilon_s = 0.0092(4)$. The rotation error, $\epsilon_{\pi}$, is bounded to be $<0.001$. Using these equations, we can correct the data for measurement errors and repeat the Maximum Likelihood Estimation (MLE) analysis to find the measurement error on the fidelity, $0.006(2)$, as given in the main text. 

% Due to the complexity of our ion state readout scheme, shown in Fig. \ref{fig:shelving}, a number of careful calibrations and simulations are needed to measure the total impact on the ion-photon entangled state fidelity. 

To determine the relevant contributions to the readout error, given above, we quantify the error due to the shelving sequence, the Raman $\pi$-pulse, and the $395\:$ms lifetime of the $4D_{5/2}$ level \cite{UDportal}. The Raman $\pi$-pulse error, $\epsilon_s$, characterization is done using a fit to Rabi oscillation data. The fit function accounts for spontaneous emission from the $5P_{3/2}$ level, which is the main source of error, as well as Gaussian decay from sources such as intensity and pointing noise. The error in the $\pi$-pulse is characterized to be $0.0092(4)$. As this is limited by spontaneous emission, it could be further improved in the future with a larger detuning from $5P_{3/2}$. Additionally, we have error from over- or under-rotation, $\epsilon_{\pi}$, due to drift in the Rabi frequency. This error is estimated to be $<0.001$ based on how often we calibrate the Raman beams during tomography. 

We use an Optical Bloch Equations simulation to reproduce the errors that occur during electron shelving and fluorescence detection, $\epsilon_b$ and $\epsilon_d$. 
The most important parameter to quantify is the polarization of the $1004\:$nm beam, since the shelving is highly sensitive to it. We use the simulation to generate a plot of the expected readout fidelity as a function of both the $1004\:$nm detuning and polarization (see Fig. \ref{fig:shelving_characterization}). 
We then use an experimental sequence that initializes the $|0\rangle$ or $|1\rangle$ state, shelves, and detects the ion state. We run this sequence as a function of $1004\:$nm detuning. We fit the initial simulation result to the resulting data in order to output the polarization error, which is measured to be $9(1)\times 10^{-4}$.

\section*{Appendix B: Characterization of Excitation Pulse Errors}

\begin{figure}
    \centering
    \includegraphics[width=1\linewidth]{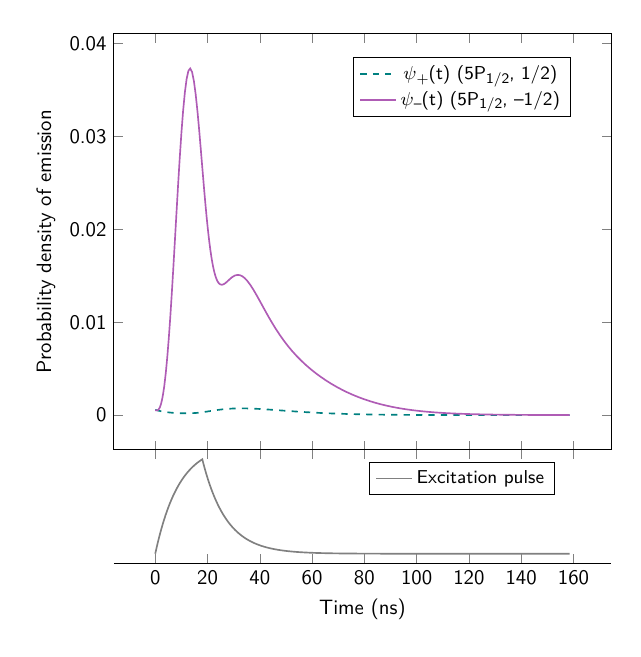}
    \caption{Probability density of 1092 nm photon emission over time. The sum of these curves integrates to unity. The purple line, $\psi_-(t)$, represents the probability density that the detected photon came from $|5P_{1/2},-1/2\rangle$ as desired. The green dashed curve, $\psi_+(t)$ represents the ion-photon experiment state preparation error, where we instead excite the ion to $|5P_{1/2},+1/2 \rangle$ before emission of a photon. The bottom, gray curve shows the profile of the excitation pulse. This is not a square pulse because the length of the pulse is comparable to the rise time of the AOM.}
    \label{fig:sigma_pulse}
\end{figure}

We perform state preparation for ion-photon entanglement using polarization sensitive optical pumping and excitation pulses. We characterize the error in these polarizations using a fit of the Optical Bloch Equations. Due to the readout method detailed in Appendix A, this error will not have a direct effect on the state fidelity, but it will cause a decrease in the entanglement generation rate. 

To find the polarization error, we execute a calibration experiment, the results of which are fit to a model of the relevant three-level system. Experimentally, we optically pump to $|5S_{1/2}, +1/2\rangle$ ($|5S_{1/2}, -1/2\rangle$) using the 422 nm $\sigma^+$ ($\sigma^-$) beam and a saturated 1092 nm repump beam. We ensure that the 1092 nm beam turns off last so no population remains in $4D_{3/2}$. We then turn on the same 422 $\sigma^+$ ($\sigma^-$) beam and record the pattern of $1092\:$nm photon arrivals as a function of time. Photons are detected due to imperfections in the polarizations of the 422 $\sigma$ beam being studied. 
We extract a polarization error of $0.016(4)$ for the $\sigma^+$ beam and $0.0088(8)$ for the $\sigma^-$ beam. 
Due to the geometry of our experimental setup, the 422 nm $\sigma$ beams are constrained to a small mirror ($\diameter1$ mm), which limits our range when aligning them to the quantization axis. Therefore, the larger polarization error compared to the $1004\:$nm beam is expected.

This first simulation is performed in the regime where the experimental parameters are constant \cite{Hugothesis}.  
We only fit to the counts that arrive after the rise time of our AOM, as the simulation is not equipped to account for these fast time dynamics. The short detection window from the ion-photon entanglement experiments is not used here, so this simulation can accurately reproduce the longer time-scale behavior. Moreover, before fitting, we normalize the data and total $5P_{1/2}$ population by their integral over the time range, which eliminates the need to know the photon detection efficiency.

We then use a second simulation to find the prepared ion-photon state with the optical pumping and excitation beams characterized. The goal is to see how population distributes across the $5P_{1/2}$ manifold as a function of the excitation pulse time, where $\psi_-(t)$ and $\psi_+(t)$ denote the amount of the total population in $|5P_{1/2},-1/2\rangle$ and $|5P_{1/2},+1/2\rangle$, respectively. The pulse duration corresponding to optimal state preparation is similar to the rise time of our excitation beam AOM, so we must include the resulting pulse shape in this simulation. Fig. \ref{fig:sigma_pulse} shows the probability density of photon emission as a function of time. In the ion-photon entanglement experiments, we limit the detection window to be within a range $t_i$ to $t_f$ after the start of the excitation pulse. The resulting ion-photon state is
\begin{equation}
\begin{aligned}
    |\psi_{\text{exp}}\rangle=\sqrt{S}\bigg(\frac{\sqrt{3}}{2}|\sigma^+\rangle|-3/2\rangle+\frac{1}{2}|\sigma^-\rangle|+1/2\rangle\bigg)
    \\
    +\sqrt{1-S}\bigg(\frac{1}{2}|\sigma^+\rangle|-1/2\rangle+\frac{\sqrt{3}}{2}|\sigma^-\rangle|+3/2\rangle\bigg),
    \end{aligned}
\end{equation}
where all ion states are defined by $|m_j\rangle$ within $D_{3/2}$, and 
\begin{equation}
    S=\int_{t_i}^{t_f}\psi_{-}(t)dt
\end{equation}
is the probability that the ion was in $|5P_{1/2},-1/2\rangle$ upon emission of the photon. The error on the ion-photon state, $1-S$, for a $3\:$ns detection window is $0.0107(4)$. Due to our detection scheme, this error will appear only as a reduction in the success probability.

The excitation pulse errors will decrease the rate by decreasing the probability of successfully generating the desired entangled state. We predict the probability of ion-photon entanglement success using the simulation with beam parameters determined as described above, as well as the measured detection efficiency of the imaging system. With a $3\:$ns detection window, we expect a success probability of $1.98(1)\times10^{-4}$ in comparison with the measured $2.07(2)\times10^{-4}$, verifying the accuracy of our analysis. We attribute the small difference between these numbers to fluctuations in the beam intensity or direction over the course of the day, as the calibration sequence and the tomography data were taken at different times. 

\section*{Appendix C: Quantum State Tomography}
We reconstruct the density matrices of the prepared entangled states using quantum state tomography, as discussed in the main text. To do so, we measure each qubit in the three Pauli bases. We use three waveplates in the photon path to enable arbitrary polarization rotation. The first $\lambda/4$ waveplate is used to rotate the incoming polarization state to be linear. The subsequent $\lambda/2$ and $\lambda/4$ allow us to perform projective measurements in $(H,D,R)$. The ion measurement is heralded upon detection of a photon, whereby the ion is then measured in $(X,Y,Z)$ using $\pi/2$ Raman pulses as applicable. We use MLE to ensure the final density matrix is physical, following the methods used in \cite{tomo_quant_inf}
with a directly-calculated raw density matrix as the input state.

\section*{Appendix D: Purity and Fidelity}

The purity sets a bound on the fidelity between the measured state and a given pure state, which gives useful information about the experimental error.
We begin with a quantum depolarizing channel on a pure state, $|\psi\rangle$,
\begin{equation}
    \rho = \lambda|\psi\rangle\langle\psi|+\frac{(1-\lambda)}{d}\mathbb{I},
\end{equation}
where $\lambda$ is the channel parameter and $d=2^n$ is the dimension for $n$ qubits.
Using $\mathbb{I}^{\otimes n}$, this equation can be written as \cite{purity}
\begin{equation}
    \rho = \left(\lambda+\frac{1-\lambda}{d}\right)|\psi\rangle\langle\psi| + 
    \left(\frac{1-\lambda}{d}\right)\sum_{i}^{d-1}
    |\psi^{\bot}_i\rangle\langle\psi^{\bot}_i|,
\end{equation}
where $|\psi^{\bot}_i\rangle$ are an orthonormal set of vectors spanning the space perpendicular to $|\psi\rangle$. In the limit that $\lambda=0$, this results in a completely mixed state.
The purity, Tr$(\rho^2)$, of this state is given by 
\begin{equation} \label{eq:purity_channel_param}
    P(\lambda)=\left(\lambda+\frac{1-\lambda}{d}\right)^2+ \left(\frac{1-\lambda}{d}\right)^2(d-1).
\end{equation}
We can relate $\lambda$ to the fidelity, $F=\langle|\psi|\rho|\psi\rangle$, by
\begin{equation}
    F(\lambda)=\frac{\lambda(d-1)+1}{d},
\end{equation}
where $d=4$ for a two-qubit state \cite{fidelity_channel_param}.
Equation \ref{eq:purity_channel_param} can now be written as a function of fidelity, which we then solve to find 
\begin{equation}
    F_{max} = \frac{1}{4}\left(1+\sqrt{3(4P-1})\right)
\end{equation}
for a two qubit state.

\section*{Appendix E: Fiber Characterization}

We perform several tests to characterize the properties of the deployed fiber. While the single-mode cutoff wavelength of the SMF-28e$+$ fiber is $1260\:$nm, the $1092\:$nm photons lie in the regime where only one additional mode exists \cite{jung2009comparative}. Tests with $1092$ nm laser light coupled into the deployed fiber show that the output exhibits no higher-order modes detectable on a beam profiler, validating that the fiber is effectively single-mode at our wavelength. Using a $3\:$km test spool of fiber within the laboratory, the attenuation of $1092\:$nm in SMF-28e$+$ was measured to be $0.77(3)$ dB/km. We measure $1092\:$nm transmission through the deployed fiber loop to be $0.43(1)$, corresponding to a loss rate of $1.31(3)$ dB/km. We attribute the heightened attenuation in the deployed fiber to three splices along its length and bend losses across the fiber, to which sub-single-mode wavelengths are more susceptible. 

\bibliographystyle{apsrev4-1}
\bibliography{biblio}{}

%merlin.mbs apsrev4-1.bst 2010-07-25 4.21a (PWD, AO, DPC) hacked
%Control: key (0)
%Control: author (72) initials jnrlst
%Control: editor formatted (1) identically to author
%Control: production of article title (-1) disabled
%Control: page (0) single
%Control: year (1) truncated
%Control: production of eprint (0) enabled
\begin{thebibliography}{58}%
\makeatletter
\providecommand \@ifxundefined [1]{%
 \@ifx{#1\undefined}
}%
\providecommand \@ifnum [1]{%
 \ifnum #1\expandafter \@firstoftwo
 \else \expandafter \@secondoftwo
 \fi
}%
\providecommand \@ifx [1]{%
 \ifx #1\expandafter \@firstoftwo
 \else \expandafter \@secondoftwo
 \fi
}%
\providecommand \natexlab [1]{#1}%
\providecommand \enquote  [1]{``#1''}%
\providecommand \bibnamefont  [1]{#1}%
\providecommand \bibfnamefont [1]{#1}%
\providecommand \citenamefont [1]{#1}%
\providecommand \href@noop [0]{\@secondoftwo}%
\providecommand \href [0]{\begingroup \@sanitize@url \@href}%
\providecommand \@href[1]{\@@startlink{#1}\@@href}%
\providecommand \@@href[1]{\endgroup#1\@@endlink}%
\providecommand \@sanitize@url [0]{\catcode `\\12\catcode `\$12\catcode `\&12\catcode `\#12\catcode `\^12\catcode `\_12\catcode `\%12\relax}%
\providecommand \@@startlink[1]{}%
\providecommand \@@endlink[0]{}%
\providecommand \url  [0]{\begingroup\@sanitize@url \@url }%
\providecommand \@url [1]{\endgroup\@href {#1}{\urlprefix }}%
\providecommand \urlprefix  [0]{URL }%
\providecommand \Eprint [0]{\href }%
\providecommand \doibase [0]{http://dx.doi.org/}%
\providecommand \selectlanguage [0]{\@gobble}%
\providecommand \bibinfo  [0]{\@secondoftwo}%
\providecommand \bibfield  [0]{\@secondoftwo}%
\providecommand \translation [1]{[#1]}%
\providecommand \BibitemOpen [0]{}%
\providecommand \bibitemStop [0]{}%
\providecommand \bibitemNoStop [0]{.\EOS\space}%
\providecommand \EOS [0]{\spacefactor3000\relax}%
\providecommand \BibitemShut  [1]{\csname bibitem#1\endcsname}%
\let\auto@bib@innerbib\@empty
%</preamble>
\bibitem [{\citenamefont {Wehner}\ \emph {et~al.}(2018)\citenamefont {Wehner}, \citenamefont {Elkouss},\ and\ \citenamefont {Hanson}}]{qinternet}%
  \BibitemOpen
  \bibfield  {author} {\bibinfo {author} {\bibfnamefont {S.}~\bibnamefont {Wehner}}, \bibinfo {author} {\bibfnamefont {D.}~\bibnamefont {Elkouss}}, \ and\ \bibinfo {author} {\bibfnamefont {R.}~\bibnamefont {Hanson}},\ }\href {\doibase 10.1126/science.aam9288} {\bibfield  {journal} {\bibinfo  {journal} {Science}\ }\textbf {\bibinfo {volume} {362}} (\bibinfo {year} {2018}),\ 10.1126/science.aam9288}\BibitemShut {NoStop}%
\bibitem [{\citenamefont {Awschalom}\ \emph {et~al.}(2021)\citenamefont {Awschalom} \emph {et~al.}}]{interconnect_overview}%
  \BibitemOpen
  \bibfield  {author} {\bibinfo {author} {\bibfnamefont {D.}~\bibnamefont {Awschalom}} \emph {et~al.},\ }\href {\doibase 10.1103/PRXQuantum.2.017002} {\bibfield  {journal} {\bibinfo  {journal} {PRX Quantum}\ }\textbf {\bibinfo {volume} {2}},\ \bibinfo {pages} {017002} (\bibinfo {year} {2021})}\BibitemShut {NoStop}%
\bibitem [{\citenamefont {Monroe}\ \emph {et~al.}(2014)\citenamefont {Monroe}, \citenamefont {Raussendorf}, \citenamefont {Ruthven}, \citenamefont {Brown}, \citenamefont {Maunz}, \citenamefont {Duan},\ and\ \citenamefont {Kim}}]{Monroe_interconnects}%
  \BibitemOpen
  \bibfield  {author} {\bibinfo {author} {\bibfnamefont {C.}~\bibnamefont {Monroe}}, \bibinfo {author} {\bibfnamefont {R.}~\bibnamefont {Raussendorf}}, \bibinfo {author} {\bibfnamefont {A.}~\bibnamefont {Ruthven}}, \bibinfo {author} {\bibfnamefont {K.~R.}\ \bibnamefont {Brown}}, \bibinfo {author} {\bibfnamefont {P.}~\bibnamefont {Maunz}}, \bibinfo {author} {\bibfnamefont {L.-M.}\ \bibnamefont {Duan}}, \ and\ \bibinfo {author} {\bibfnamefont {J.}~\bibnamefont {Kim}},\ }\href {\doibase 10.1103/PhysRevA.89.022317} {\bibfield  {journal} {\bibinfo  {journal} {Phys. Rev. A}\ }\textbf {\bibinfo {volume} {89}},\ \bibinfo {pages} {022317} (\bibinfo {year} {2014})}\BibitemShut {NoStop}%
\bibitem [{\citenamefont {Sinclair}\ \emph {et~al.}(2025)\citenamefont {Sinclair}, \citenamefont {Ramette}, \citenamefont {Grinkemeyer}, \citenamefont {Bluvstein}, \citenamefont {Lukin},\ and\ \citenamefont {Vuleti\ifmmode~\acute{c}\else \'{c}\fi{}}}]{neutral_interconnects_overview}%
  \BibitemOpen
  \bibfield  {author} {\bibinfo {author} {\bibfnamefont {J.}~\bibnamefont {Sinclair}}, \bibinfo {author} {\bibfnamefont {J.}~\bibnamefont {Ramette}}, \bibinfo {author} {\bibfnamefont {B.}~\bibnamefont {Grinkemeyer}}, \bibinfo {author} {\bibfnamefont {D.}~\bibnamefont {Bluvstein}}, \bibinfo {author} {\bibfnamefont {M.~D.}\ \bibnamefont {Lukin}}, \ and\ \bibinfo {author} {\bibfnamefont {V.}~\bibnamefont {Vuleti\ifmmode~\acute{c}\else \'{c}\fi{}}},\ }\href {\doibase 10.1103/PhysRevResearch.7.013313} {\bibfield  {journal} {\bibinfo  {journal} {Phys. Rev. Res.}\ }\textbf {\bibinfo {volume} {7}},\ \bibinfo {pages} {013313} (\bibinfo {year} {2025})}\BibitemShut {NoStop}%
\bibitem [{\citenamefont {Covey}\ \emph {et~al.}(2023)\citenamefont {Covey}, \citenamefont {Weinfurter},\ and\ \citenamefont {Bernien}}]{Covey2023_review}%
  \BibitemOpen
  \bibfield  {author} {\bibinfo {author} {\bibfnamefont {J.~P.}\ \bibnamefont {Covey}}, \bibinfo {author} {\bibfnamefont {H.}~\bibnamefont {Weinfurter}}, \ and\ \bibinfo {author} {\bibfnamefont {H.}~\bibnamefont {Bernien}},\ }\href {\doibase 10.1038/s41534-023-00759-9} {\bibfield  {journal} {\bibinfo  {journal} {npj Quantum Information}\ }\textbf {\bibinfo {volume} {9}},\ \bibinfo {pages} {90} (\bibinfo {year} {2023})}\BibitemShut {NoStop}%
\bibitem [{\citenamefont {Gottesman}\ \emph {et~al.}(2012)\citenamefont {Gottesman}, \citenamefont {Jennewein},\ and\ \citenamefont {Croke}}]{telescope_repeaters}%
  \BibitemOpen
  \bibfield  {author} {\bibinfo {author} {\bibfnamefont {D.}~\bibnamefont {Gottesman}}, \bibinfo {author} {\bibfnamefont {T.}~\bibnamefont {Jennewein}}, \ and\ \bibinfo {author} {\bibfnamefont {S.}~\bibnamefont {Croke}},\ }\href@noop {} {\bibfield  {journal} {\bibinfo  {journal} {Phys. Rev. Lett.}\ }\textbf {\bibinfo {volume} {109}},\ \bibinfo {pages} {070503} (\bibinfo {year} {2012})}\BibitemShut {NoStop}%
\bibitem [{\citenamefont {Nichol}\ \emph {et~al.}(2022)\citenamefont {Nichol}, \citenamefont {Srinivas}, \citenamefont {Nadlinger}, \citenamefont {Drmota}, \citenamefont {Main}, \citenamefont {Araneda}, \citenamefont {Ballance},\ and\ \citenamefont {Lucas}}]{Nichol2022}%
  \BibitemOpen
  \bibfield  {author} {\bibinfo {author} {\bibfnamefont {B.~C.}\ \bibnamefont {Nichol}}, \bibinfo {author} {\bibfnamefont {R.}~\bibnamefont {Srinivas}}, \bibinfo {author} {\bibfnamefont {D.~P.}\ \bibnamefont {Nadlinger}}, \bibinfo {author} {\bibfnamefont {P.}~\bibnamefont {Drmota}}, \bibinfo {author} {\bibfnamefont {D.}~\bibnamefont {Main}}, \bibinfo {author} {\bibfnamefont {G.}~\bibnamefont {Araneda}}, \bibinfo {author} {\bibfnamefont {C.~J.}\ \bibnamefont {Ballance}}, \ and\ \bibinfo {author} {\bibfnamefont {D.~M.}\ \bibnamefont {Lucas}},\ }\href {\doibase 10.1038/s41586-022-05088-z} {\bibfield  {journal} {\bibinfo  {journal} {Nature}\ }\textbf {\bibinfo {volume} {609}},\ \bibinfo {pages} {689} (\bibinfo {year} {2022})}\BibitemShut {NoStop}%
\bibitem [{\citenamefont {Drmota}\ \emph {et~al.}(2024)\citenamefont {Drmota}, \citenamefont {Nadlinger}, \citenamefont {Main}, \citenamefont {Nichol}, \citenamefont {Ainley}, \citenamefont {Leichtle}, \citenamefont {Mantri}, \citenamefont {Kashefi}, \citenamefont {Srinivas}, \citenamefont {Araneda}, \citenamefont {Ballance},\ and\ \citenamefont {Lucas}}]{Drmota_blindQC}%
  \BibitemOpen
  \bibfield  {author} {\bibinfo {author} {\bibfnamefont {P.}~\bibnamefont {Drmota}}, \bibinfo {author} {\bibfnamefont {D.~P.}\ \bibnamefont {Nadlinger}}, \bibinfo {author} {\bibfnamefont {D.}~\bibnamefont {Main}}, \bibinfo {author} {\bibfnamefont {B.~C.}\ \bibnamefont {Nichol}}, \bibinfo {author} {\bibfnamefont {E.~M.}\ \bibnamefont {Ainley}}, \bibinfo {author} {\bibfnamefont {D.}~\bibnamefont {Leichtle}}, \bibinfo {author} {\bibfnamefont {A.}~\bibnamefont {Mantri}}, \bibinfo {author} {\bibfnamefont {E.}~\bibnamefont {Kashefi}}, \bibinfo {author} {\bibfnamefont {R.}~\bibnamefont {Srinivas}}, \bibinfo {author} {\bibfnamefont {G.}~\bibnamefont {Araneda}}, \bibinfo {author} {\bibfnamefont {C.~J.}\ \bibnamefont {Ballance}}, \ and\ \bibinfo {author} {\bibfnamefont {D.~M.}\ \bibnamefont {Lucas}},\ }\href {\doibase 10.1103/PhysRevLett.132.150604} {\bibfield  {journal} {\bibinfo  {journal} {Phys. Rev. Lett.}\ }\textbf {\bibinfo {volume} {132}},\ \bibinfo {pages} {150604} (\bibinfo {year} {2024})}\BibitemShut
  {NoStop}%
\bibitem [{\citenamefont {Fitzsimons}(2017)}]{Fitzsimons2017_blindQC_overview}%
  \BibitemOpen
  \bibfield  {author} {\bibinfo {author} {\bibfnamefont {J.~F.}\ \bibnamefont {Fitzsimons}},\ }\href {\doibase 10.1038/s41534-017-0025-3} {\bibfield  {journal} {\bibinfo  {journal} {npj Quantum Information}\ }\textbf {\bibinfo {volume} {3}},\ \bibinfo {pages} {23} (\bibinfo {year} {2017})}\BibitemShut {NoStop}%
\bibitem [{\citenamefont {Gisin}\ \emph {et~al.}(2002)\citenamefont {Gisin}, \citenamefont {Ribordy}, \citenamefont {Tittel},\ and\ \citenamefont {Zbinden}}]{crypto}%
  \BibitemOpen
  \bibfield  {author} {\bibinfo {author} {\bibfnamefont {N.}~\bibnamefont {Gisin}}, \bibinfo {author} {\bibfnamefont {G.}~\bibnamefont {Ribordy}}, \bibinfo {author} {\bibfnamefont {W.}~\bibnamefont {Tittel}}, \ and\ \bibinfo {author} {\bibfnamefont {H.}~\bibnamefont {Zbinden}},\ }\href@noop {} {\bibfield  {journal} {\bibinfo  {journal} {Rev. Mod. Phys.}\ }\textbf {\bibinfo {volume} {74}},\ \bibinfo {pages} {145} (\bibinfo {year} {2002})}\BibitemShut {NoStop}%
\bibitem [{\citenamefont {Ac\'{\i}n}\ \emph {et~al.}(2007)\citenamefont {Ac\'{\i}n}, \citenamefont {Brunner}, \citenamefont {Gisin}, \citenamefont {Massar}, \citenamefont {Pironio},\ and\ \citenamefont {Scarani}}]{diqkd_theory}%
  \BibitemOpen
  \bibfield  {author} {\bibinfo {author} {\bibfnamefont {A.}~\bibnamefont {Ac\'{\i}n}}, \bibinfo {author} {\bibfnamefont {N.}~\bibnamefont {Brunner}}, \bibinfo {author} {\bibfnamefont {N.}~\bibnamefont {Gisin}}, \bibinfo {author} {\bibfnamefont {S.}~\bibnamefont {Massar}}, \bibinfo {author} {\bibfnamefont {S.}~\bibnamefont {Pironio}}, \ and\ \bibinfo {author} {\bibfnamefont {V.}~\bibnamefont {Scarani}},\ }\href {\doibase 10.1103/PhysRevLett.98.230501} {\bibfield  {journal} {\bibinfo  {journal} {Phys. Rev. Lett.}\ }\textbf {\bibinfo {volume} {98}},\ \bibinfo {pages} {230501} (\bibinfo {year} {2007})}\BibitemShut {NoStop}%
\bibitem [{\citenamefont {Nadlinger}\ \emph {et~al.}(2022)\citenamefont {Nadlinger}, \citenamefont {Drmota}, \citenamefont {Nichol}, \citenamefont {Araneda}, \citenamefont {Main}, \citenamefont {Srinivas}, \citenamefont {Lucas}, \citenamefont {Ballance}, \citenamefont {Ivanov}, \citenamefont {Tan}, \citenamefont {Sekatski}, \citenamefont {Urbanke}, \citenamefont {Renner}, \citenamefont {Sangouard},\ and\ \citenamefont {Bancal}}]{Nadlinger2022}%
  \BibitemOpen
  \bibfield  {author} {\bibinfo {author} {\bibfnamefont {D.~P.}\ \bibnamefont {Nadlinger}}, \bibinfo {author} {\bibfnamefont {P.}~\bibnamefont {Drmota}}, \bibinfo {author} {\bibfnamefont {B.~C.}\ \bibnamefont {Nichol}}, \bibinfo {author} {\bibfnamefont {G.}~\bibnamefont {Araneda}}, \bibinfo {author} {\bibfnamefont {D.}~\bibnamefont {Main}}, \bibinfo {author} {\bibfnamefont {R.}~\bibnamefont {Srinivas}}, \bibinfo {author} {\bibfnamefont {D.~M.}\ \bibnamefont {Lucas}}, \bibinfo {author} {\bibfnamefont {C.~J.}\ \bibnamefont {Ballance}}, \bibinfo {author} {\bibfnamefont {K.}~\bibnamefont {Ivanov}}, \bibinfo {author} {\bibfnamefont {E.~Y.-Z.}\ \bibnamefont {Tan}}, \bibinfo {author} {\bibfnamefont {P.}~\bibnamefont {Sekatski}}, \bibinfo {author} {\bibfnamefont {R.~L.}\ \bibnamefont {Urbanke}}, \bibinfo {author} {\bibfnamefont {R.}~\bibnamefont {Renner}}, \bibinfo {author} {\bibfnamefont {N.}~\bibnamefont {Sangouard}}, \ and\ \bibinfo {author} {\bibfnamefont {J.-D.}\ \bibnamefont {Bancal}},\ }\href@noop {}
  {\bibfield  {journal} {\bibinfo  {journal} {Nature}\ }\textbf {\bibinfo {volume} {607}},\ \bibinfo {pages} {682} (\bibinfo {year} {2022})}\BibitemShut {NoStop}%
\bibitem [{\citenamefont {Zhang}\ \emph {et~al.}(2022)\citenamefont {Zhang}, \citenamefont {van Leent}, \citenamefont {Redeker}, \citenamefont {Garthoff}, \citenamefont {Schwonnek}, \citenamefont {Fertig}, \citenamefont {Eppelt}, \citenamefont {Rosenfeld}, \citenamefont {Scarani}, \citenamefont {Lim},\ and\ \citenamefont {Weinfurter}}]{Zhang2022_diqkd}%
  \BibitemOpen
  \bibfield  {author} {\bibinfo {author} {\bibfnamefont {W.}~\bibnamefont {Zhang}}, \bibinfo {author} {\bibfnamefont {T.}~\bibnamefont {van Leent}}, \bibinfo {author} {\bibfnamefont {K.}~\bibnamefont {Redeker}}, \bibinfo {author} {\bibfnamefont {R.}~\bibnamefont {Garthoff}}, \bibinfo {author} {\bibfnamefont {R.}~\bibnamefont {Schwonnek}}, \bibinfo {author} {\bibfnamefont {F.}~\bibnamefont {Fertig}}, \bibinfo {author} {\bibfnamefont {S.}~\bibnamefont {Eppelt}}, \bibinfo {author} {\bibfnamefont {W.}~\bibnamefont {Rosenfeld}}, \bibinfo {author} {\bibfnamefont {V.}~\bibnamefont {Scarani}}, \bibinfo {author} {\bibfnamefont {C.~C.-W.}\ \bibnamefont {Lim}}, \ and\ \bibinfo {author} {\bibfnamefont {H.}~\bibnamefont {Weinfurter}},\ }\href@noop {} {\bibfield  {journal} {\bibinfo  {journal} {Nature}\ }\textbf {\bibinfo {volume} {607}},\ \bibinfo {pages} {687} (\bibinfo {year} {2022})}\BibitemShut {NoStop}%
\bibitem [{\citenamefont {Hensen}\ \emph {et~al.}(2015)\citenamefont {Hensen}, \citenamefont {Bernien}, \citenamefont {Dr{\'e}au}, \citenamefont {Reiserer}, \citenamefont {Kalb}, \citenamefont {Blok}, \citenamefont {Ruitenberg}, \citenamefont {Vermeulen}, \citenamefont {Schouten}, \citenamefont {Abell{\'a}n}, \citenamefont {Amaya}, \citenamefont {Pruneri}, \citenamefont {Mitchell}, \citenamefont {Markham}, \citenamefont {Twitchen}, \citenamefont {Elkouss}, \citenamefont {Wehner}, \citenamefont {Taminiau},\ and\ \citenamefont {Hanson}}]{Hensen2015}%
  \BibitemOpen
  \bibfield  {author} {\bibinfo {author} {\bibfnamefont {B.}~\bibnamefont {Hensen}}, \bibinfo {author} {\bibfnamefont {H.}~\bibnamefont {Bernien}}, \bibinfo {author} {\bibfnamefont {A.~E.}\ \bibnamefont {Dr{\'e}au}}, \bibinfo {author} {\bibfnamefont {A.}~\bibnamefont {Reiserer}}, \bibinfo {author} {\bibfnamefont {N.}~\bibnamefont {Kalb}}, \bibinfo {author} {\bibfnamefont {M.~S.}\ \bibnamefont {Blok}}, \bibinfo {author} {\bibfnamefont {J.}~\bibnamefont {Ruitenberg}}, \bibinfo {author} {\bibfnamefont {R.~F.~L.}\ \bibnamefont {Vermeulen}}, \bibinfo {author} {\bibfnamefont {R.~N.}\ \bibnamefont {Schouten}}, \bibinfo {author} {\bibfnamefont {C.}~\bibnamefont {Abell{\'a}n}}, \bibinfo {author} {\bibfnamefont {W.}~\bibnamefont {Amaya}}, \bibinfo {author} {\bibfnamefont {V.}~\bibnamefont {Pruneri}}, \bibinfo {author} {\bibfnamefont {M.~W.}\ \bibnamefont {Mitchell}}, \bibinfo {author} {\bibfnamefont {M.}~\bibnamefont {Markham}}, \bibinfo {author} {\bibfnamefont {D.~J.}\ \bibnamefont {Twitchen}}, \bibinfo {author}
  {\bibfnamefont {D.}~\bibnamefont {Elkouss}}, \bibinfo {author} {\bibfnamefont {S.}~\bibnamefont {Wehner}}, \bibinfo {author} {\bibfnamefont {T.~H.}\ \bibnamefont {Taminiau}}, \ and\ \bibinfo {author} {\bibfnamefont {R.}~\bibnamefont {Hanson}},\ }\href {\doibase 10.1038/nature15759} {\bibfield  {journal} {\bibinfo  {journal} {Nature}\ }\textbf {\bibinfo {volume} {526}},\ \bibinfo {pages} {682} (\bibinfo {year} {2015})}\BibitemShut {NoStop}%
\bibitem [{\citenamefont {Rosenfeld}\ \emph {et~al.}(2017)\citenamefont {Rosenfeld}, \citenamefont {Burchardt}, \citenamefont {Garthoff}, \citenamefont {Redeker}, \citenamefont {Ortegel}, \citenamefont {Rau},\ and\ \citenamefont {Weinfurter}}]{bell_test_atoms}%
  \BibitemOpen
  \bibfield  {author} {\bibinfo {author} {\bibfnamefont {W.}~\bibnamefont {Rosenfeld}}, \bibinfo {author} {\bibfnamefont {D.}~\bibnamefont {Burchardt}}, \bibinfo {author} {\bibfnamefont {R.}~\bibnamefont {Garthoff}}, \bibinfo {author} {\bibfnamefont {K.}~\bibnamefont {Redeker}}, \bibinfo {author} {\bibfnamefont {N.}~\bibnamefont {Ortegel}}, \bibinfo {author} {\bibfnamefont {M.}~\bibnamefont {Rau}}, \ and\ \bibinfo {author} {\bibfnamefont {H.}~\bibnamefont {Weinfurter}},\ }\href {\doibase 10.1103/PhysRevLett.119.010402} {\bibfield  {journal} {\bibinfo  {journal} {Phys. Rev. Lett.}\ }\textbf {\bibinfo {volume} {119}},\ \bibinfo {pages} {010402} (\bibinfo {year} {2017})}\BibitemShut {NoStop}%
\bibitem [{\citenamefont {Krutyanskiy}\ \emph {et~al.}(2024)\citenamefont {Krutyanskiy}, \citenamefont {Canteri}, \citenamefont {Meraner}, \citenamefont {Krcmarsky},\ and\ \citenamefont {Lanyon}}]{inns2024_101km}%
  \BibitemOpen
  \bibfield  {author} {\bibinfo {author} {\bibfnamefont {V.}~\bibnamefont {Krutyanskiy}}, \bibinfo {author} {\bibfnamefont {M.}~\bibnamefont {Canteri}}, \bibinfo {author} {\bibfnamefont {M.}~\bibnamefont {Meraner}}, \bibinfo {author} {\bibfnamefont {V.}~\bibnamefont {Krcmarsky}}, \ and\ \bibinfo {author} {\bibfnamefont {B.}~\bibnamefont {Lanyon}},\ }\href {\doibase 10.1103/PRXQuantum.5.020308} {\bibfield  {journal} {\bibinfo  {journal} {PRX Quantum}\ }\textbf {\bibinfo {volume} {5}},\ \bibinfo {pages} {020308} (\bibinfo {year} {2024})}\BibitemShut {NoStop}%
\bibitem [{\citenamefont {Kucera}\ \emph {et~al.}(2024)\citenamefont {Kucera}, \citenamefont {Haen}, \citenamefont {Arensk{\"o}tter}, \citenamefont {Bauer}, \citenamefont {Meiers}, \citenamefont {Sch{\"a}fer}, \citenamefont {Boland}, \citenamefont {Yahyapour}, \citenamefont {Lessing}, \citenamefont {Holzwarth}, \citenamefont {Becher},\ and\ \citenamefont {Eschner}}]{Kucera2024_14km_network}%
  \BibitemOpen
  \bibfield  {author} {\bibinfo {author} {\bibfnamefont {S.}~\bibnamefont {Kucera}}, \bibinfo {author} {\bibfnamefont {C.}~\bibnamefont {Haen}}, \bibinfo {author} {\bibfnamefont {E.}~\bibnamefont {Arensk{\"o}tter}}, \bibinfo {author} {\bibfnamefont {T.}~\bibnamefont {Bauer}}, \bibinfo {author} {\bibfnamefont {J.}~\bibnamefont {Meiers}}, \bibinfo {author} {\bibfnamefont {M.}~\bibnamefont {Sch{\"a}fer}}, \bibinfo {author} {\bibfnamefont {R.}~\bibnamefont {Boland}}, \bibinfo {author} {\bibfnamefont {M.}~\bibnamefont {Yahyapour}}, \bibinfo {author} {\bibfnamefont {M.}~\bibnamefont {Lessing}}, \bibinfo {author} {\bibfnamefont {R.}~\bibnamefont {Holzwarth}}, \bibinfo {author} {\bibfnamefont {C.}~\bibnamefont {Becher}}, \ and\ \bibinfo {author} {\bibfnamefont {J.}~\bibnamefont {Eschner}},\ }\href {\doibase 10.1038/s41534-024-00886-x} {\bibfield  {journal} {\bibinfo  {journal} {npj Quantum Information}\ }\textbf {\bibinfo {volume} {10}},\ \bibinfo {pages} {88} (\bibinfo {year} {2024})}\BibitemShut {NoStop}%
\bibitem [{\citenamefont {Ritter}\ \emph {et~al.}(2012)\citenamefont {Ritter}, \citenamefont {N{\"o}lleke}, \citenamefont {Hahn}, \citenamefont {Reiserer}, \citenamefont {Neuzner}, \citenamefont {Uphoff}, \citenamefont {M{\"u}cke}, \citenamefont {Figueroa}, \citenamefont {Bochmann},\ and\ \citenamefont {Rempe}}]{Ritter2012}%
  \BibitemOpen
  \bibfield  {author} {\bibinfo {author} {\bibfnamefont {S.}~\bibnamefont {Ritter}}, \bibinfo {author} {\bibfnamefont {C.}~\bibnamefont {N{\"o}lleke}}, \bibinfo {author} {\bibfnamefont {C.}~\bibnamefont {Hahn}}, \bibinfo {author} {\bibfnamefont {A.}~\bibnamefont {Reiserer}}, \bibinfo {author} {\bibfnamefont {A.}~\bibnamefont {Neuzner}}, \bibinfo {author} {\bibfnamefont {M.}~\bibnamefont {Uphoff}}, \bibinfo {author} {\bibfnamefont {M.}~\bibnamefont {M{\"u}cke}}, \bibinfo {author} {\bibfnamefont {E.}~\bibnamefont {Figueroa}}, \bibinfo {author} {\bibfnamefont {J.}~\bibnamefont {Bochmann}}, \ and\ \bibinfo {author} {\bibfnamefont {G.}~\bibnamefont {Rempe}},\ }\href {\doibase 10.1038/nature11023} {\bibfield  {journal} {\bibinfo  {journal} {Nature}\ }\textbf {\bibinfo {volume} {484}},\ \bibinfo {pages} {195} (\bibinfo {year} {2012})}\BibitemShut {NoStop}%
\bibitem [{\citenamefont {van Leent}\ \emph {et~al.}(2022)\citenamefont {van Leent}, \citenamefont {Bock}, \citenamefont {Fertig}, \citenamefont {Garthoff}, \citenamefont {Eppelt}, \citenamefont {Zhou}, \citenamefont {Malik}, \citenamefont {Seubert}, \citenamefont {Bauer}, \citenamefont {Rosenfeld}, \citenamefont {Zhang}, \citenamefont {Becher},\ and\ \citenamefont {Weinfurter}}]{vanLeent2022_neutrals_33km}%
  \BibitemOpen
  \bibfield  {author} {\bibinfo {author} {\bibfnamefont {T.}~\bibnamefont {van Leent}}, \bibinfo {author} {\bibfnamefont {M.}~\bibnamefont {Bock}}, \bibinfo {author} {\bibfnamefont {F.}~\bibnamefont {Fertig}}, \bibinfo {author} {\bibfnamefont {R.}~\bibnamefont {Garthoff}}, \bibinfo {author} {\bibfnamefont {S.}~\bibnamefont {Eppelt}}, \bibinfo {author} {\bibfnamefont {Y.}~\bibnamefont {Zhou}}, \bibinfo {author} {\bibfnamefont {P.}~\bibnamefont {Malik}}, \bibinfo {author} {\bibfnamefont {M.}~\bibnamefont {Seubert}}, \bibinfo {author} {\bibfnamefont {T.}~\bibnamefont {Bauer}}, \bibinfo {author} {\bibfnamefont {W.}~\bibnamefont {Rosenfeld}}, \bibinfo {author} {\bibfnamefont {W.}~\bibnamefont {Zhang}}, \bibinfo {author} {\bibfnamefont {C.}~\bibnamefont {Becher}}, \ and\ \bibinfo {author} {\bibfnamefont {H.}~\bibnamefont {Weinfurter}},\ }\href {\doibase 10.1038/s41586-022-04764-4} {\bibfield  {journal} {\bibinfo  {journal} {Nature}\ }\textbf {\bibinfo {volume} {607}},\ \bibinfo {pages} {69} (\bibinfo {year}
  {2022})}\BibitemShut {NoStop}%
\bibitem [{\citenamefont {Stolk}\ \emph {et~al.}(2024)\citenamefont {Stolk}, \citenamefont {van~der Enden}, \citenamefont {Slater}, \citenamefont {te~Raa-Derckx}, \citenamefont {Botma}, \citenamefont {van Rantwijk}, \citenamefont {Biemond}, \citenamefont {Hagen}, \citenamefont {Herfst}, \citenamefont {Koek}, \citenamefont {Meskers}, \citenamefont {Vollmer}, \citenamefont {van Zwet}, \citenamefont {Markham}, \citenamefont {Edmonds}, \citenamefont {Geus}, \citenamefont {Elsen}, \citenamefont {Jungbluth}, \citenamefont {Haefner}, \citenamefont {Tresp}, \citenamefont {Stuhler}, \citenamefont {Ritter},\ and\ \citenamefont {Hanson}}]{stolk_dist_classical_network}%
  \BibitemOpen
  \bibfield  {author} {\bibinfo {author} {\bibfnamefont {A.~J.}\ \bibnamefont {Stolk}}, \bibinfo {author} {\bibfnamefont {K.~L.}\ \bibnamefont {van~der Enden}}, \bibinfo {author} {\bibfnamefont {M.-C.}\ \bibnamefont {Slater}}, \bibinfo {author} {\bibfnamefont {I.}~\bibnamefont {te~Raa-Derckx}}, \bibinfo {author} {\bibfnamefont {P.}~\bibnamefont {Botma}}, \bibinfo {author} {\bibfnamefont {J.}~\bibnamefont {van Rantwijk}}, \bibinfo {author} {\bibfnamefont {J.~J.~B.}\ \bibnamefont {Biemond}}, \bibinfo {author} {\bibfnamefont {R.~A.~J.}\ \bibnamefont {Hagen}}, \bibinfo {author} {\bibfnamefont {R.~W.}\ \bibnamefont {Herfst}}, \bibinfo {author} {\bibfnamefont {W.~D.}\ \bibnamefont {Koek}}, \bibinfo {author} {\bibfnamefont {A.~J.~H.}\ \bibnamefont {Meskers}}, \bibinfo {author} {\bibfnamefont {R.}~\bibnamefont {Vollmer}}, \bibinfo {author} {\bibfnamefont {E.~J.}\ \bibnamefont {van Zwet}}, \bibinfo {author} {\bibfnamefont {M.}~\bibnamefont {Markham}}, \bibinfo {author} {\bibfnamefont {A.~M.}\ \bibnamefont {Edmonds}},
  \bibinfo {author} {\bibfnamefont {J.~F.}\ \bibnamefont {Geus}}, \bibinfo {author} {\bibfnamefont {F.}~\bibnamefont {Elsen}}, \bibinfo {author} {\bibfnamefont {B.}~\bibnamefont {Jungbluth}}, \bibinfo {author} {\bibfnamefont {C.}~\bibnamefont {Haefner}}, \bibinfo {author} {\bibfnamefont {C.}~\bibnamefont {Tresp}}, \bibinfo {author} {\bibfnamefont {J.}~\bibnamefont {Stuhler}}, \bibinfo {author} {\bibfnamefont {S.}~\bibnamefont {Ritter}}, \ and\ \bibinfo {author} {\bibfnamefont {R.}~\bibnamefont {Hanson}},\ }\href@noop {} {\bibfield  {journal} {\bibinfo  {journal} {Science Advances}\ }\textbf {\bibinfo {volume} {10}} (\bibinfo {year} {2024})}\BibitemShut {NoStop}%
\bibitem [{\citenamefont {Knaut}\ \emph {et~al.}(2024)\citenamefont {Knaut}, \citenamefont {Suleymanzade}, \citenamefont {Wei}, \citenamefont {Assumpcao}, \citenamefont {Stas}, \citenamefont {Huan}, \citenamefont {Machielse}, \citenamefont {Knall}, \citenamefont {Sutula}, \citenamefont {Baranes}, \citenamefont {Sinclair}, \citenamefont {De-Eknamkul}, \citenamefont {Levonian}, \citenamefont {Bhaskar}, \citenamefont {Park}, \citenamefont {Lon{\v{c}}ar},\ and\ \citenamefont {Lukin}}]{Knaut2024_bostonnetwork}%
  \BibitemOpen
  \bibfield  {author} {\bibinfo {author} {\bibfnamefont {C.~M.}\ \bibnamefont {Knaut}}, \bibinfo {author} {\bibfnamefont {A.}~\bibnamefont {Suleymanzade}}, \bibinfo {author} {\bibfnamefont {Y.-C.}\ \bibnamefont {Wei}}, \bibinfo {author} {\bibfnamefont {D.~R.}\ \bibnamefont {Assumpcao}}, \bibinfo {author} {\bibfnamefont {P.-J.}\ \bibnamefont {Stas}}, \bibinfo {author} {\bibfnamefont {Y.~Q.}\ \bibnamefont {Huan}}, \bibinfo {author} {\bibfnamefont {B.}~\bibnamefont {Machielse}}, \bibinfo {author} {\bibfnamefont {E.~N.}\ \bibnamefont {Knall}}, \bibinfo {author} {\bibfnamefont {M.}~\bibnamefont {Sutula}}, \bibinfo {author} {\bibfnamefont {G.}~\bibnamefont {Baranes}}, \bibinfo {author} {\bibfnamefont {N.}~\bibnamefont {Sinclair}}, \bibinfo {author} {\bibfnamefont {C.}~\bibnamefont {De-Eknamkul}}, \bibinfo {author} {\bibfnamefont {D.~S.}\ \bibnamefont {Levonian}}, \bibinfo {author} {\bibfnamefont {M.~K.}\ \bibnamefont {Bhaskar}}, \bibinfo {author} {\bibfnamefont {H.}~\bibnamefont {Park}}, \bibinfo {author}
  {\bibfnamefont {M.}~\bibnamefont {Lon{\v{c}}ar}}, \ and\ \bibinfo {author} {\bibfnamefont {M.~D.}\ \bibnamefont {Lukin}},\ }\href {\doibase 10.1038/s41586-024-07252-z} {\bibfield  {journal} {\bibinfo  {journal} {Nature}\ }\textbf {\bibinfo {volume} {629}},\ \bibinfo {pages} {573} (\bibinfo {year} {2024})}\BibitemShut {NoStop}%
\bibitem [{\citenamefont {Stockill}\ \emph {et~al.}(2017)\citenamefont {Stockill}, \citenamefont {Stanley}, \citenamefont {Huthmacher}, \citenamefont {Clarke}, \citenamefont {Hugues}, \citenamefont {Miller}, \citenamefont {Matthiesen}, \citenamefont {Le~Gall},\ and\ \citenamefont {Atat\"ure}}]{quantumdots}%
  \BibitemOpen
  \bibfield  {author} {\bibinfo {author} {\bibfnamefont {R.}~\bibnamefont {Stockill}}, \bibinfo {author} {\bibfnamefont {M.~J.}\ \bibnamefont {Stanley}}, \bibinfo {author} {\bibfnamefont {L.}~\bibnamefont {Huthmacher}}, \bibinfo {author} {\bibfnamefont {E.}~\bibnamefont {Clarke}}, \bibinfo {author} {\bibfnamefont {M.}~\bibnamefont {Hugues}}, \bibinfo {author} {\bibfnamefont {A.~J.}\ \bibnamefont {Miller}}, \bibinfo {author} {\bibfnamefont {C.}~\bibnamefont {Matthiesen}}, \bibinfo {author} {\bibfnamefont {C.}~\bibnamefont {Le~Gall}}, \ and\ \bibinfo {author} {\bibfnamefont {M.}~\bibnamefont {Atat\"ure}},\ }\href@noop {} {\bibfield  {journal} {\bibinfo  {journal} {Phys. Rev. Lett.}\ }\textbf {\bibinfo {volume} {119}},\ \bibinfo {pages} {010503} (\bibinfo {year} {2017})}\BibitemShut {NoStop}%
\bibitem [{\citenamefont {Wang}\ \emph {et~al.}(2021)\citenamefont {Wang}, \citenamefont {Luan}, \citenamefont {Qiao}, \citenamefont {Um}, \citenamefont {Zhang}, \citenamefont {Wang}, \citenamefont {Yuan}, \citenamefont {Gu}, \citenamefont {Zhang},\ and\ \citenamefont {Kim}}]{longest_coherence}%
  \BibitemOpen
  \bibfield  {author} {\bibinfo {author} {\bibfnamefont {P.}~\bibnamefont {Wang}}, \bibinfo {author} {\bibfnamefont {C.-Y.}\ \bibnamefont {Luan}}, \bibinfo {author} {\bibfnamefont {M.}~\bibnamefont {Qiao}}, \bibinfo {author} {\bibfnamefont {M.}~\bibnamefont {Um}}, \bibinfo {author} {\bibfnamefont {J.}~\bibnamefont {Zhang}}, \bibinfo {author} {\bibfnamefont {Y.}~\bibnamefont {Wang}}, \bibinfo {author} {\bibfnamefont {X.}~\bibnamefont {Yuan}}, \bibinfo {author} {\bibfnamefont {M.}~\bibnamefont {Gu}}, \bibinfo {author} {\bibfnamefont {J.}~\bibnamefont {Zhang}}, \ and\ \bibinfo {author} {\bibfnamefont {K.}~\bibnamefont {Kim}},\ }\href {\doibase 10.1038/s41467-020-20330-w} {\bibfield  {journal} {\bibinfo  {journal} {Nature Communications}\ }\textbf {\bibinfo {volume} {12}},\ \bibinfo {pages} {233} (\bibinfo {year} {2021})}\BibitemShut {NoStop}%
\bibitem [{\citenamefont {Smith}\ \emph {et~al.}(2025)\citenamefont {Smith}, \citenamefont {Leu}, \citenamefont {Miyanishi}, \citenamefont {Gely},\ and\ \citenamefont {Lucas}}]{smith2025_single_qubit_uwave}%
  \BibitemOpen
  \bibfield  {author} {\bibinfo {author} {\bibfnamefont {M.~C.}\ \bibnamefont {Smith}}, \bibinfo {author} {\bibfnamefont {A.~D.}\ \bibnamefont {Leu}}, \bibinfo {author} {\bibfnamefont {K.}~\bibnamefont {Miyanishi}}, \bibinfo {author} {\bibfnamefont {M.~F.}\ \bibnamefont {Gely}}, \ and\ \bibinfo {author} {\bibfnamefont {D.~M.}\ \bibnamefont {Lucas}},\ }\href {https://arxiv.org/abs/2412.04421} {} (\bibinfo {year} {2025}),\ \Eprint {http://arxiv.org/abs/2412.04421} {arXiv:2412.04421 [quant-ph]} \BibitemShut {NoStop}%
\bibitem [{\citenamefont {Löschnauer}\ \emph {et~al.}(2024)\citenamefont {Löschnauer}, \citenamefont {Toba}, \citenamefont {Hughes}, \citenamefont {King}, \citenamefont {Weber}, \citenamefont {Srinivas}, \citenamefont {Matt}, \citenamefont {Nourshargh}, \citenamefont {Allcock}, \citenamefont {Ballance}, \citenamefont {Matthiesen}, \citenamefont {Malinowski},\ and\ \citenamefont {Harty}}]{löschnauer2024_best2qubit_ion}%
  \BibitemOpen
  \bibfield  {author} {\bibinfo {author} {\bibfnamefont {C.~M.}\ \bibnamefont {Löschnauer}}, \bibinfo {author} {\bibfnamefont {J.~M.}\ \bibnamefont {Toba}}, \bibinfo {author} {\bibfnamefont {A.~C.}\ \bibnamefont {Hughes}}, \bibinfo {author} {\bibfnamefont {S.~A.}\ \bibnamefont {King}}, \bibinfo {author} {\bibfnamefont {M.~A.}\ \bibnamefont {Weber}}, \bibinfo {author} {\bibfnamefont {R.}~\bibnamefont {Srinivas}}, \bibinfo {author} {\bibfnamefont {R.}~\bibnamefont {Matt}}, \bibinfo {author} {\bibfnamefont {R.}~\bibnamefont {Nourshargh}}, \bibinfo {author} {\bibfnamefont {D.~T.~C.}\ \bibnamefont {Allcock}}, \bibinfo {author} {\bibfnamefont {C.~J.}\ \bibnamefont {Ballance}}, \bibinfo {author} {\bibfnamefont {C.}~\bibnamefont {Matthiesen}}, \bibinfo {author} {\bibfnamefont {M.}~\bibnamefont {Malinowski}}, \ and\ \bibinfo {author} {\bibfnamefont {T.~P.}\ \bibnamefont {Harty}},\ }\href {https://arxiv.org/abs/2407.07694} {} (\bibinfo {year} {2024}),\ \Eprint {http://arxiv.org/abs/2407.07694} {arXiv:2407.07694
  [quant-ph]} \BibitemShut {NoStop}%
\bibitem [{\citenamefont {Main}\ \emph {et~al.}(2025)\citenamefont {Main}, \citenamefont {Drmota}, \citenamefont {Nadlinger}, \citenamefont {Ainley}, \citenamefont {Agrawal}, \citenamefont {Nichol}, \citenamefont {Srinivas}, \citenamefont {Araneda},\ and\ \citenamefont {Lucas}}]{Main2025}%
  \BibitemOpen
  \bibfield  {author} {\bibinfo {author} {\bibfnamefont {D.}~\bibnamefont {Main}}, \bibinfo {author} {\bibfnamefont {P.}~\bibnamefont {Drmota}}, \bibinfo {author} {\bibfnamefont {D.~P.}\ \bibnamefont {Nadlinger}}, \bibinfo {author} {\bibfnamefont {E.~M.}\ \bibnamefont {Ainley}}, \bibinfo {author} {\bibfnamefont {A.}~\bibnamefont {Agrawal}}, \bibinfo {author} {\bibfnamefont {B.~C.}\ \bibnamefont {Nichol}}, \bibinfo {author} {\bibfnamefont {R.}~\bibnamefont {Srinivas}}, \bibinfo {author} {\bibfnamefont {G.}~\bibnamefont {Araneda}}, \ and\ \bibinfo {author} {\bibfnamefont {D.~M.}\ \bibnamefont {Lucas}},\ }\href {\doibase 10.1038/s41586-024-08404-x} {\bibfield  {journal} {\bibinfo  {journal} {Nature}\ }\textbf {\bibinfo {volume} {638}},\ \bibinfo {pages} {383} (\bibinfo {year} {2025})}\BibitemShut {NoStop}%
\bibitem [{\citenamefont {Saha}\ \emph {et~al.}(2025)\citenamefont {Saha}, \citenamefont {Shalaev}, \citenamefont {O'Reilly}, \citenamefont {Goetting}, \citenamefont {Toh}, \citenamefont {Kalakuntla}, \citenamefont {Yu},\ and\ \citenamefont {Monroe}}]{Saha2025}%
  \BibitemOpen
  \bibfield  {author} {\bibinfo {author} {\bibfnamefont {S.}~\bibnamefont {Saha}}, \bibinfo {author} {\bibfnamefont {M.}~\bibnamefont {Shalaev}}, \bibinfo {author} {\bibfnamefont {J.}~\bibnamefont {O'Reilly}}, \bibinfo {author} {\bibfnamefont {I.}~\bibnamefont {Goetting}}, \bibinfo {author} {\bibfnamefont {G.}~\bibnamefont {Toh}}, \bibinfo {author} {\bibfnamefont {A.}~\bibnamefont {Kalakuntla}}, \bibinfo {author} {\bibfnamefont {Y.}~\bibnamefont {Yu}}, \ and\ \bibinfo {author} {\bibfnamefont {C.}~\bibnamefont {Monroe}},\ }\href {\doibase 10.1038/s41467-025-57557-4} {\bibfield  {journal} {\bibinfo  {journal} {Nature Communications}\ }\textbf {\bibinfo {volume} {16}},\ \bibinfo {pages} {2533} (\bibinfo {year} {2025})}\BibitemShut {NoStop}%
\bibitem [{\citenamefont {O'Reilly}\ \emph {et~al.}(2024)\citenamefont {O'Reilly}, \citenamefont {Toh}, \citenamefont {Goetting}, \citenamefont {Saha}, \citenamefont {Shalaev}, \citenamefont {Carter}, \citenamefont {Risinger}, \citenamefont {Kalakuntla}, \citenamefont {Li}, \citenamefont {Verma},\ and\ \citenamefont {Monroe}}]{Oreilly_symp_cool}%
  \BibitemOpen
  \bibfield  {author} {\bibinfo {author} {\bibfnamefont {J.}~\bibnamefont {O'Reilly}}, \bibinfo {author} {\bibfnamefont {G.}~\bibnamefont {Toh}}, \bibinfo {author} {\bibfnamefont {I.}~\bibnamefont {Goetting}}, \bibinfo {author} {\bibfnamefont {S.}~\bibnamefont {Saha}}, \bibinfo {author} {\bibfnamefont {M.}~\bibnamefont {Shalaev}}, \bibinfo {author} {\bibfnamefont {A.~L.}\ \bibnamefont {Carter}}, \bibinfo {author} {\bibfnamefont {A.}~\bibnamefont {Risinger}}, \bibinfo {author} {\bibfnamefont {A.}~\bibnamefont {Kalakuntla}}, \bibinfo {author} {\bibfnamefont {T.}~\bibnamefont {Li}}, \bibinfo {author} {\bibfnamefont {A.}~\bibnamefont {Verma}}, \ and\ \bibinfo {author} {\bibfnamefont {C.}~\bibnamefont {Monroe}},\ }\href {\doibase 10.1103/PhysRevLett.133.090802} {\bibfield  {journal} {\bibinfo  {journal} {Phys. Rev. Lett.}\ }\textbf {\bibinfo {volume} {133}},\ \bibinfo {pages} {090802} (\bibinfo {year} {2024})}\BibitemShut {NoStop}%
\bibitem [{\citenamefont {Krutyanskiy}\ \emph {et~al.}(2023)\citenamefont {Krutyanskiy}, \citenamefont {Galli}, \citenamefont {Krcmarsky}, \citenamefont {Baier}, \citenamefont {Fioretto}, \citenamefont {Pu}, \citenamefont {Mazloom}, \citenamefont {Sekatski}, \citenamefont {Canteri}, \citenamefont {Teller}, \citenamefont {Schupp}, \citenamefont {Bate}, \citenamefont {Meraner}, \citenamefont {Sangouard}, \citenamefont {Lanyon},\ and\ \citenamefont {Northup}}]{inns_remote_ent}%
  \BibitemOpen
  \bibfield  {author} {\bibinfo {author} {\bibfnamefont {V.}~\bibnamefont {Krutyanskiy}}, \bibinfo {author} {\bibfnamefont {M.}~\bibnamefont {Galli}}, \bibinfo {author} {\bibfnamefont {V.}~\bibnamefont {Krcmarsky}}, \bibinfo {author} {\bibfnamefont {S.}~\bibnamefont {Baier}}, \bibinfo {author} {\bibfnamefont {D.~A.}\ \bibnamefont {Fioretto}}, \bibinfo {author} {\bibfnamefont {Y.}~\bibnamefont {Pu}}, \bibinfo {author} {\bibfnamefont {A.}~\bibnamefont {Mazloom}}, \bibinfo {author} {\bibfnamefont {P.}~\bibnamefont {Sekatski}}, \bibinfo {author} {\bibfnamefont {M.}~\bibnamefont {Canteri}}, \bibinfo {author} {\bibfnamefont {M.}~\bibnamefont {Teller}}, \bibinfo {author} {\bibfnamefont {J.}~\bibnamefont {Schupp}}, \bibinfo {author} {\bibfnamefont {J.}~\bibnamefont {Bate}}, \bibinfo {author} {\bibfnamefont {M.}~\bibnamefont {Meraner}}, \bibinfo {author} {\bibfnamefont {N.}~\bibnamefont {Sangouard}}, \bibinfo {author} {\bibfnamefont {B.~P.}\ \bibnamefont {Lanyon}}, \ and\ \bibinfo {author} {\bibfnamefont {T.~E.}\
  \bibnamefont {Northup}},\ }\href {\doibase 10.1103/PhysRevLett.130.050803} {\bibfield  {journal} {\bibinfo  {journal} {Phys. Rev. Lett.}\ }\textbf {\bibinfo {volume} {130}},\ \bibinfo {pages} {050803} (\bibinfo {year} {2023})}\BibitemShut {NoStop}%
\bibitem [{\citenamefont {Ikuta}\ \emph {et~al.}(2011)\citenamefont {Ikuta}, \citenamefont {Kusaka}, \citenamefont {Kitano}, \citenamefont {Kato}, \citenamefont {Yamamoto}, \citenamefont {Koashi},\ and\ \citenamefont {Imoto}}]{Ikuta2011}%
  \BibitemOpen
  \bibfield  {author} {\bibinfo {author} {\bibfnamefont {R.}~\bibnamefont {Ikuta}}, \bibinfo {author} {\bibfnamefont {Y.}~\bibnamefont {Kusaka}}, \bibinfo {author} {\bibfnamefont {T.}~\bibnamefont {Kitano}}, \bibinfo {author} {\bibfnamefont {H.}~\bibnamefont {Kato}}, \bibinfo {author} {\bibfnamefont {T.}~\bibnamefont {Yamamoto}}, \bibinfo {author} {\bibfnamefont {M.}~\bibnamefont {Koashi}}, \ and\ \bibinfo {author} {\bibfnamefont {N.}~\bibnamefont {Imoto}},\ }\href@noop {} {\bibfield  {journal} {\bibinfo  {journal} {Nature Communications}\ }\textbf {\bibinfo {volume} {2}},\ \bibinfo {pages} {537} (\bibinfo {year} {2011})}\BibitemShut {NoStop}%
\bibitem [{\citenamefont {Geus}\ \emph {et~al.}(2024)\citenamefont {Geus}, \citenamefont {Elsen}, \citenamefont {Nyga}, \citenamefont {Stolk}, \citenamefont {van~der Enden}, \citenamefont {van Zwet}, \citenamefont {Haefner}, \citenamefont {Hanson},\ and\ \citenamefont {Jungbluth}}]{Geus24_hansonQFC}%
  \BibitemOpen
  \bibfield  {author} {\bibinfo {author} {\bibfnamefont {J.~F.}\ \bibnamefont {Geus}}, \bibinfo {author} {\bibfnamefont {F.}~\bibnamefont {Elsen}}, \bibinfo {author} {\bibfnamefont {S.}~\bibnamefont {Nyga}}, \bibinfo {author} {\bibfnamefont {A.~J.}\ \bibnamefont {Stolk}}, \bibinfo {author} {\bibfnamefont {K.~L.}\ \bibnamefont {van~der Enden}}, \bibinfo {author} {\bibfnamefont {E.~J.}\ \bibnamefont {van Zwet}}, \bibinfo {author} {\bibfnamefont {C.}~\bibnamefont {Haefner}}, \bibinfo {author} {\bibfnamefont {R.}~\bibnamefont {Hanson}}, \ and\ \bibinfo {author} {\bibfnamefont {B.}~\bibnamefont {Jungbluth}},\ }\href {\doibase 10.1364/OPTICAQ.515769} {\bibfield  {journal} {\bibinfo  {journal} {Optica Quantum}\ }\textbf {\bibinfo {volume} {2}},\ \bibinfo {pages} {189} (\bibinfo {year} {2024})}\BibitemShut {NoStop}%
\bibitem [{\citenamefont {Bock}\ \emph {et~al.}(2018)\citenamefont {Bock}, \citenamefont {Eich}, \citenamefont {Kucera}, \citenamefont {Kreis}, \citenamefont {Lenhard}, \citenamefont {Becher},\ and\ \citenamefont {Eschner}}]{Bock2018}%
  \BibitemOpen
  \bibfield  {author} {\bibinfo {author} {\bibfnamefont {M.}~\bibnamefont {Bock}}, \bibinfo {author} {\bibfnamefont {P.}~\bibnamefont {Eich}}, \bibinfo {author} {\bibfnamefont {S.}~\bibnamefont {Kucera}}, \bibinfo {author} {\bibfnamefont {M.}~\bibnamefont {Kreis}}, \bibinfo {author} {\bibfnamefont {A.}~\bibnamefont {Lenhard}}, \bibinfo {author} {\bibfnamefont {C.}~\bibnamefont {Becher}}, \ and\ \bibinfo {author} {\bibfnamefont {J.}~\bibnamefont {Eschner}},\ }\href {\doibase 10.1038/s41467-018-04341-2} {\bibfield  {journal} {\bibinfo  {journal} {Nature Communications}\ }\textbf {\bibinfo {volume} {9}},\ \bibinfo {pages} {1998} (\bibinfo {year} {2018})}\BibitemShut {NoStop}%
\bibitem [{\citenamefont {Saha}\ \emph {et~al.}(2023)\citenamefont {Saha}, \citenamefont {Siverns}, \citenamefont {Hannegan}, \citenamefont {Quraishi},\ and\ \citenamefont {Waks}}]{Quraishi2023_QFC}%
  \BibitemOpen
  \bibfield  {author} {\bibinfo {author} {\bibfnamefont {U.}~\bibnamefont {Saha}}, \bibinfo {author} {\bibfnamefont {J.~D.}\ \bibnamefont {Siverns}}, \bibinfo {author} {\bibfnamefont {J.}~\bibnamefont {Hannegan}}, \bibinfo {author} {\bibfnamefont {Q.}~\bibnamefont {Quraishi}}, \ and\ \bibinfo {author} {\bibfnamefont {E.}~\bibnamefont {Waks}},\ }\href {\doibase 10.1021/acsphotonics.3c00581} {\bibfield  {journal} {\bibinfo  {journal} {ACS Photonics}\ }\textbf {\bibinfo {volume} {10}},\ \bibinfo {pages} {2861} (\bibinfo {year} {2023})}\BibitemShut {NoStop}%
\bibitem [{\citenamefont {Tchebotareva}\ \emph {et~al.}(2019)\citenamefont {Tchebotareva}, \citenamefont {Hermans}, \citenamefont {Humphreys}, \citenamefont {Voigt}, \citenamefont {Harmsma}, \citenamefont {Cheng}, \citenamefont {Verlaan}, \citenamefont {Dijkhuizen}, \citenamefont {de~Jong}, \citenamefont {Dr\'eau},\ and\ \citenamefont {Hanson}}]{QFC_eff_17}%
  \BibitemOpen
  \bibfield  {author} {\bibinfo {author} {\bibfnamefont {A.}~\bibnamefont {Tchebotareva}}, \bibinfo {author} {\bibfnamefont {S.~L.~N.}\ \bibnamefont {Hermans}}, \bibinfo {author} {\bibfnamefont {P.~C.}\ \bibnamefont {Humphreys}}, \bibinfo {author} {\bibfnamefont {D.}~\bibnamefont {Voigt}}, \bibinfo {author} {\bibfnamefont {P.~J.}\ \bibnamefont {Harmsma}}, \bibinfo {author} {\bibfnamefont {L.~K.}\ \bibnamefont {Cheng}}, \bibinfo {author} {\bibfnamefont {A.~L.}\ \bibnamefont {Verlaan}}, \bibinfo {author} {\bibfnamefont {N.}~\bibnamefont {Dijkhuizen}}, \bibinfo {author} {\bibfnamefont {W.}~\bibnamefont {de~Jong}}, \bibinfo {author} {\bibfnamefont {A.}~\bibnamefont {Dr\'eau}}, \ and\ \bibinfo {author} {\bibfnamefont {R.}~\bibnamefont {Hanson}},\ }\href {\doibase 10.1103/PhysRevLett.123.063601} {\bibfield  {journal} {\bibinfo  {journal} {Phys. Rev. Lett.}\ }\textbf {\bibinfo {volume} {123}},\ \bibinfo {pages} {063601} (\bibinfo {year} {2019})}\BibitemShut {NoStop}%
\bibitem [{\citenamefont {Krutyanskiy}\ \emph {et~al.}(2019)\citenamefont {Krutyanskiy}, \citenamefont {Meraner}, \citenamefont {Schupp}, \citenamefont {Krcmarsky}, \citenamefont {Hainzer},\ and\ \citenamefont {Lanyon}}]{QFC_25}%
  \BibitemOpen
  \bibfield  {author} {\bibinfo {author} {\bibfnamefont {V.}~\bibnamefont {Krutyanskiy}}, \bibinfo {author} {\bibfnamefont {M.}~\bibnamefont {Meraner}}, \bibinfo {author} {\bibfnamefont {J.}~\bibnamefont {Schupp}}, \bibinfo {author} {\bibfnamefont {V.}~\bibnamefont {Krcmarsky}}, \bibinfo {author} {\bibfnamefont {H.}~\bibnamefont {Hainzer}}, \ and\ \bibinfo {author} {\bibfnamefont {B.~P.}\ \bibnamefont {Lanyon}},\ }\href@noop {} {\bibfield  {journal} {\bibinfo  {journal} {npj Quantum Information}\ }\textbf {\bibinfo {volume} {5}},\ \bibinfo {pages} {72} (\bibinfo {year} {2019})}\BibitemShut {NoStop}%
\bibitem [{\citenamefont {Luo}\ \emph {et~al.}(2022)\citenamefont {Luo}, \citenamefont {Yu}, \citenamefont {Liu}, \citenamefont {Zheng}, \citenamefont {Wang}, \citenamefont {Wang}, \citenamefont {Li}, \citenamefont {Jiang}, \citenamefont {Xie}, \citenamefont {Zhang}, \citenamefont {Bao},\ and\ \citenamefont {Pan}}]{QFC_eff_46}%
  \BibitemOpen
  \bibfield  {author} {\bibinfo {author} {\bibfnamefont {X.-Y.}\ \bibnamefont {Luo}}, \bibinfo {author} {\bibfnamefont {Y.}~\bibnamefont {Yu}}, \bibinfo {author} {\bibfnamefont {J.-L.}\ \bibnamefont {Liu}}, \bibinfo {author} {\bibfnamefont {M.-Y.}\ \bibnamefont {Zheng}}, \bibinfo {author} {\bibfnamefont {C.-Y.}\ \bibnamefont {Wang}}, \bibinfo {author} {\bibfnamefont {B.}~\bibnamefont {Wang}}, \bibinfo {author} {\bibfnamefont {J.}~\bibnamefont {Li}}, \bibinfo {author} {\bibfnamefont {X.}~\bibnamefont {Jiang}}, \bibinfo {author} {\bibfnamefont {X.-P.}\ \bibnamefont {Xie}}, \bibinfo {author} {\bibfnamefont {Q.}~\bibnamefont {Zhang}}, \bibinfo {author} {\bibfnamefont {X.-H.}\ \bibnamefont {Bao}}, \ and\ \bibinfo {author} {\bibfnamefont {J.-W.}\ \bibnamefont {Pan}},\ }\href {\doibase 10.1103/PhysRevLett.129.050503} {\bibfield  {journal} {\bibinfo  {journal} {Phys. Rev. Lett.}\ }\textbf {\bibinfo {volume} {129}},\ \bibinfo {pages} {050503} (\bibinfo {year} {2022})}\BibitemShut {NoStop}%
\bibitem [{\citenamefont {van Leent}\ \emph {et~al.}(2020)\citenamefont {van Leent}, \citenamefont {Bock}, \citenamefont {Garthoff}, \citenamefont {Redeker}, \citenamefont {Zhang}, \citenamefont {Bauer}, \citenamefont {Rosenfeld}, \citenamefont {Becher},\ and\ \citenamefont {Weinfurter}}]{QFC_eff_57}%
  \BibitemOpen
  \bibfield  {author} {\bibinfo {author} {\bibfnamefont {T.}~\bibnamefont {van Leent}}, \bibinfo {author} {\bibfnamefont {M.}~\bibnamefont {Bock}}, \bibinfo {author} {\bibfnamefont {R.}~\bibnamefont {Garthoff}}, \bibinfo {author} {\bibfnamefont {K.}~\bibnamefont {Redeker}}, \bibinfo {author} {\bibfnamefont {W.}~\bibnamefont {Zhang}}, \bibinfo {author} {\bibfnamefont {T.}~\bibnamefont {Bauer}}, \bibinfo {author} {\bibfnamefont {W.}~\bibnamefont {Rosenfeld}}, \bibinfo {author} {\bibfnamefont {C.}~\bibnamefont {Becher}}, \ and\ \bibinfo {author} {\bibfnamefont {H.}~\bibnamefont {Weinfurter}},\ }\href {\doibase 10.1103/PhysRevLett.124.010510} {\bibfield  {journal} {\bibinfo  {journal} {Phys. Rev. Lett.}\ }\textbf {\bibinfo {volume} {124}},\ \bibinfo {pages} {010510} (\bibinfo {year} {2020})}\BibitemShut {NoStop}%
\bibitem [{\citenamefont {Li}\ \emph {et~al.}(2025)\citenamefont {Li}, \citenamefont {Hu}, \citenamefont {Jia}, \citenamefont {Huie}, \citenamefont {Sun}, \citenamefont {{Aakash}}, \citenamefont {Dong}, \citenamefont {Hiri-O-Tuppa},\ and\ \citenamefont {Covey}}]{covey_telecom_network}%
  \BibitemOpen
  \bibfield  {author} {\bibinfo {author} {\bibfnamefont {L.}~\bibnamefont {Li}}, \bibinfo {author} {\bibfnamefont {X.}~\bibnamefont {Hu}}, \bibinfo {author} {\bibfnamefont {Z.}~\bibnamefont {Jia}}, \bibinfo {author} {\bibfnamefont {W.}~\bibnamefont {Huie}}, \bibinfo {author} {\bibfnamefont {W.~K.~C.}\ \bibnamefont {Sun}}, \bibinfo {author} {\bibnamefont {{Aakash}}}, \bibinfo {author} {\bibfnamefont {Y.}~\bibnamefont {Dong}}, \bibinfo {author} {\bibfnamefont {N.}~\bibnamefont {Hiri-O-Tuppa}}, \ and\ \bibinfo {author} {\bibfnamefont {J.~P.}\ \bibnamefont {Covey}},\ }\href@noop {} {\bibfield  {journal} {\bibinfo  {journal} {Nature Physics}\ } (\bibinfo {year} {2025})}\BibitemShut {NoStop}%
\bibitem [{\citenamefont {Uysal}\ \emph {et~al.}(2025)\citenamefont {Uysal}, \citenamefont {Dusanowski}, \citenamefont {Xu}, \citenamefont {Horvath}, \citenamefont {Ourari}, \citenamefont {Cava}, \citenamefont {de~Leon},\ and\ \citenamefont {Thompson}}]{erbium}%
  \BibitemOpen
  \bibfield  {author} {\bibinfo {author} {\bibfnamefont {M.~T.}\ \bibnamefont {Uysal}}, \bibinfo {author} {\bibfnamefont {L.}~\bibnamefont {Dusanowski}}, \bibinfo {author} {\bibfnamefont {H.}~\bibnamefont {Xu}}, \bibinfo {author} {\bibfnamefont {S.~P.}\ \bibnamefont {Horvath}}, \bibinfo {author} {\bibfnamefont {S.}~\bibnamefont {Ourari}}, \bibinfo {author} {\bibfnamefont {R.~J.}\ \bibnamefont {Cava}}, \bibinfo {author} {\bibfnamefont {N.~P.}\ \bibnamefont {de~Leon}}, \ and\ \bibinfo {author} {\bibfnamefont {J.~D.}\ \bibnamefont {Thompson}},\ }\href {\doibase 10.1103/PhysRevX.15.011071} {\bibfield  {journal} {\bibinfo  {journal} {Phys. Rev. X}\ }\textbf {\bibinfo {volume} {15}},\ \bibinfo {pages} {011071} (\bibinfo {year} {2025})}\BibitemShut {NoStop}%
\bibitem [{\citenamefont {Laccotripes}\ \emph {et~al.}(2024)\citenamefont {Laccotripes}, \citenamefont {M{\"u}ller}, \citenamefont {Stevenson}, \citenamefont {Skiba-Szymanska}, \citenamefont {Ritchie},\ and\ \citenamefont {Shields}}]{direct_spin}%
  \BibitemOpen
  \bibfield  {author} {\bibinfo {author} {\bibfnamefont {P.}~\bibnamefont {Laccotripes}}, \bibinfo {author} {\bibfnamefont {T.}~\bibnamefont {M{\"u}ller}}, \bibinfo {author} {\bibfnamefont {R.~M.}\ \bibnamefont {Stevenson}}, \bibinfo {author} {\bibfnamefont {J.}~\bibnamefont {Skiba-Szymanska}}, \bibinfo {author} {\bibfnamefont {D.~A.}\ \bibnamefont {Ritchie}}, \ and\ \bibinfo {author} {\bibfnamefont {A.~J.}\ \bibnamefont {Shields}},\ }\href {\doibase 10.1038/s41467-024-53964-1} {\bibfield  {journal} {\bibinfo  {journal} {Nature Communications}\ }\textbf {\bibinfo {volume} {15}},\ \bibinfo {pages} {9740} (\bibinfo {year} {2024})}\BibitemShut {NoStop}%
\bibitem [{\citenamefont {Schubert}(2006)}]{telecom}%
  \BibitemOpen
  \bibfield  {author} {\bibinfo {author} {\bibfnamefont {E.~F.}\ \bibnamefont {Schubert}},\ }\href@noop {} {\emph {\bibinfo {title} {Light-Emitting Diodes}}}\ (\bibinfo  {publisher} {Cambridge University Press},\ \bibinfo {year} {2006})\BibitemShut {NoStop}%
\bibitem [{\citenamefont {Barakhshan}\ \emph {et~al.}()\citenamefont {Barakhshan}, \citenamefont {Marrs}, \citenamefont {Bhosale}, \citenamefont {Arora}, \citenamefont {Eigenmann},\ and\ \citenamefont {Safronova}}]{UDportal}%
  \BibitemOpen
  \bibfield  {author} {\bibinfo {author} {\bibfnamefont {P.}~\bibnamefont {Barakhshan}}, \bibinfo {author} {\bibfnamefont {A.}~\bibnamefont {Marrs}}, \bibinfo {author} {\bibfnamefont {A.}~\bibnamefont {Bhosale}}, \bibinfo {author} {\bibfnamefont {B.}~\bibnamefont {Arora}}, \bibinfo {author} {\bibfnamefont {R.}~\bibnamefont {Eigenmann}}, \ and\ \bibinfo {author} {\bibfnamefont {M.~S.}\ \bibnamefont {Safronova}},\ }\href@noop {} {}\bibinfo {howpublished} {{ extit{Portal for High-Precision Atomic Data and Computation}} (version 2.0). University of Delaware, Newark, DE, USA. URL: {https://www.udel.edu/atom}}\BibitemShut {NoStop}%
\bibitem [{\citenamefont {Luo}\ \emph {et~al.}(2009)\citenamefont {Luo}, \citenamefont {Hayes}, \citenamefont {Manning}, \citenamefont {Matsukevich}, \citenamefont {Maunz}, \citenamefont {Olmschenk}, \citenamefont {Sterk},\ and\ \citenamefont {Monroe}}]{luo2009protocols}%
  \BibitemOpen
  \bibfield  {author} {\bibinfo {author} {\bibfnamefont {L.}~\bibnamefont {Luo}}, \bibinfo {author} {\bibfnamefont {D.}~\bibnamefont {Hayes}}, \bibinfo {author} {\bibfnamefont {T.}~\bibnamefont {Manning}}, \bibinfo {author} {\bibfnamefont {D.}~\bibnamefont {Matsukevich}}, \bibinfo {author} {\bibfnamefont {P.}~\bibnamefont {Maunz}}, \bibinfo {author} {\bibfnamefont {S.}~\bibnamefont {Olmschenk}}, \bibinfo {author} {\bibfnamefont {J.}~\bibnamefont {Sterk}}, \ and\ \bibinfo {author} {\bibfnamefont {C.}~\bibnamefont {Monroe}},\ }\href@noop {} {\bibfield  {journal} {\bibinfo  {journal} {Fortschritte der Physik}\ }\textbf {\bibinfo {volume} {57}},\ \bibinfo {pages} {1133} (\bibinfo {year} {2009})}\BibitemShut {NoStop}%
\bibitem [{\citenamefont {Simon}\ and\ \citenamefont {Irvine}(2003)}]{type_2_theory}%
  \BibitemOpen
  \bibfield  {author} {\bibinfo {author} {\bibfnamefont {C.}~\bibnamefont {Simon}}\ and\ \bibinfo {author} {\bibfnamefont {W.~T.~M.}\ \bibnamefont {Irvine}},\ }\href {\doibase 10.1103/PhysRevLett.91.110405} {\bibfield  {journal} {\bibinfo  {journal} {Phys. Rev. Lett.}\ }\textbf {\bibinfo {volume} {91}},\ \bibinfo {pages} {110405} (\bibinfo {year} {2003})}\BibitemShut {NoStop}%
\bibitem [{\citenamefont {Stute}\ \emph {et~al.}(2012)\citenamefont {Stute}, \citenamefont {Casabone}, \citenamefont {Schindler}, \citenamefont {Monz}, \citenamefont {Schmidt}, \citenamefont {Brandst{\"a}tter}, \citenamefont {Northup},\ and\ \citenamefont {Blatt}}]{Stute2012}%
  \BibitemOpen
  \bibfield  {author} {\bibinfo {author} {\bibfnamefont {A.}~\bibnamefont {Stute}}, \bibinfo {author} {\bibfnamefont {B.}~\bibnamefont {Casabone}}, \bibinfo {author} {\bibfnamefont {P.}~\bibnamefont {Schindler}}, \bibinfo {author} {\bibfnamefont {T.}~\bibnamefont {Monz}}, \bibinfo {author} {\bibfnamefont {P.~O.}\ \bibnamefont {Schmidt}}, \bibinfo {author} {\bibfnamefont {B.}~\bibnamefont {Brandst{\"a}tter}}, \bibinfo {author} {\bibfnamefont {T.~E.}\ \bibnamefont {Northup}}, \ and\ \bibinfo {author} {\bibfnamefont {R.}~\bibnamefont {Blatt}},\ }\href {\doibase 10.1038/nature11120} {\bibfield  {journal} {\bibinfo  {journal} {Nature}\ }\textbf {\bibinfo {volume} {485}},\ \bibinfo {pages} {482} (\bibinfo {year} {2012})}\BibitemShut {NoStop}%
\bibitem [{\citenamefont {Ruskuc}\ \emph {et~al.}(2025)\citenamefont {Ruskuc}, \citenamefont {Wu}, \citenamefont {Green}, \citenamefont {Hermans}, \citenamefont {Pajak}, \citenamefont {Choi},\ and\ \citenamefont {Faraon}}]{Ruskuc2025_multiplexing}%
  \BibitemOpen
  \bibfield  {author} {\bibinfo {author} {\bibfnamefont {A.}~\bibnamefont {Ruskuc}}, \bibinfo {author} {\bibfnamefont {C.-J.}\ \bibnamefont {Wu}}, \bibinfo {author} {\bibfnamefont {E.}~\bibnamefont {Green}}, \bibinfo {author} {\bibfnamefont {S.~L.~N.}\ \bibnamefont {Hermans}}, \bibinfo {author} {\bibfnamefont {W.}~\bibnamefont {Pajak}}, \bibinfo {author} {\bibfnamefont {J.}~\bibnamefont {Choi}}, \ and\ \bibinfo {author} {\bibfnamefont {A.}~\bibnamefont {Faraon}},\ }\href {\doibase 10.1038/s41586-024-08537-z} {\bibfield  {journal} {\bibinfo  {journal} {Nature}\ }\textbf {\bibinfo {volume} {639}},\ \bibinfo {pages} {54} (\bibinfo {year} {2025})}\BibitemShut {NoStop}%
\bibitem [{\citenamefont {Cui}\ \emph {et~al.}(2025)\citenamefont {Cui}, \citenamefont {Wang}, \citenamefont {Lai}, \citenamefont {Wang}, \citenamefont {Shi}, \citenamefont {Liu}, \citenamefont {Sun}, \citenamefont {Tian}, \citenamefont {Liang}, \citenamefont {Qi}, \citenamefont {Huang}, \citenamefont {Zhou}, \citenamefont {Wu}, \citenamefont {Xu}, \citenamefont {Duan},\ and\ \citenamefont {Pu}}]{duan_multiplex}%
  \BibitemOpen
  \bibfield  {author} {\bibinfo {author} {\bibfnamefont {Z.~B.}\ \bibnamefont {Cui}}, \bibinfo {author} {\bibfnamefont {Z.~Q.}\ \bibnamefont {Wang}}, \bibinfo {author} {\bibfnamefont {P.~C.}\ \bibnamefont {Lai}}, \bibinfo {author} {\bibfnamefont {Y.}~\bibnamefont {Wang}}, \bibinfo {author} {\bibfnamefont {J.~X.}\ \bibnamefont {Shi}}, \bibinfo {author} {\bibfnamefont {P.~Y.}\ \bibnamefont {Liu}}, \bibinfo {author} {\bibfnamefont {Y.~D.}\ \bibnamefont {Sun}}, \bibinfo {author} {\bibfnamefont {Z.~C.}\ \bibnamefont {Tian}}, \bibinfo {author} {\bibfnamefont {Y.~B.}\ \bibnamefont {Liang}}, \bibinfo {author} {\bibfnamefont {B.~X.}\ \bibnamefont {Qi}}, \bibinfo {author} {\bibfnamefont {Y.~Y.}\ \bibnamefont {Huang}}, \bibinfo {author} {\bibfnamefont {Z.~C.}\ \bibnamefont {Zhou}}, \bibinfo {author} {\bibfnamefont {Y.~K.}\ \bibnamefont {Wu}}, \bibinfo {author} {\bibfnamefont {Y.}~\bibnamefont {Xu}}, \bibinfo {author} {\bibfnamefont {L.~M.}\ \bibnamefont {Duan}}, \ and\ \bibinfo {author} {\bibfnamefont {Y.~F.}\
  \bibnamefont {Pu}},\ }\href {https://arxiv.org/abs/2503.13898} {} (\bibinfo {year} {2025}),\ \Eprint {http://arxiv.org/abs/2503.13898} {arXiv:2503.13898 [quant-ph]} \BibitemShut {NoStop}%
\bibitem [{\citenamefont {Je\ifmmode~\check{z}\else \v{z}\fi{}ek}\ \emph {et~al.}(2003)\citenamefont {Je\ifmmode~\check{z}\else \v{z}\fi{}ek}, \citenamefont {Fiur\'a\ifmmode~\check{s}\else \v{s}\fi{}ek},\ and\ \citenamefont {Hradil}}]{tomo_quant_inf}%
  \BibitemOpen
  \bibfield  {author} {\bibinfo {author} {\bibfnamefont {M.}~\bibnamefont {Je\ifmmode~\check{z}\else \v{z}\fi{}ek}}, \bibinfo {author} {\bibfnamefont {J.}~\bibnamefont {Fiur\'a\ifmmode~\check{s}\else \v{s}\fi{}ek}}, \ and\ \bibinfo {author} {\bibfnamefont {Z.}~\bibnamefont {Hradil}},\ }\href {\doibase 10.1103/PhysRevA.68.012305} {\bibfield  {journal} {\bibinfo  {journal} {Phys. Rev. A}\ }\textbf {\bibinfo {volume} {68}},\ \bibinfo {pages} {012305} (\bibinfo {year} {2003})}\BibitemShut {NoStop}%
\bibitem [{\citenamefont {Altepeter}\ \emph {et~al.}(2015)\citenamefont {Altepeter}, \citenamefont {Jeffrey},\ and\ \citenamefont {Kwiat}}]{tomography}%
  \BibitemOpen
  \bibfield  {author} {\bibinfo {author} {\bibfnamefont {J.}~\bibnamefont {Altepeter}}, \bibinfo {author} {\bibfnamefont {E.}~\bibnamefont {Jeffrey}}, \ and\ \bibinfo {author} {\bibfnamefont {P.}~\bibnamefont {Kwiat}}\ }(\bibinfo  {publisher} {Academic Press},\ \bibinfo {year} {2015})\ pp.\ \bibinfo {pages} {105--159}\BibitemShut {NoStop}%
\bibitem [{\citenamefont {Bussi{\`e}res}\ \emph {et~al.}(2014)\citenamefont {Bussi{\`e}res}, \citenamefont {Clausen}, \citenamefont {Tiranov}, \citenamefont {Korzh}, \citenamefont {Verma}, \citenamefont {Nam}, \citenamefont {Marsili}, \citenamefont {Ferrier}, \citenamefont {Goldner}, \citenamefont {Herrmann}, \citenamefont {Silberhorn}, \citenamefont {Sohler}, \citenamefont {Afzelius},\ and\ \citenamefont {Gisin}}]{purity}%
  \BibitemOpen
  \bibfield  {author} {\bibinfo {author} {\bibfnamefont {F.}~\bibnamefont {Bussi{\`e}res}}, \bibinfo {author} {\bibfnamefont {C.}~\bibnamefont {Clausen}}, \bibinfo {author} {\bibfnamefont {A.}~\bibnamefont {Tiranov}}, \bibinfo {author} {\bibfnamefont {B.}~\bibnamefont {Korzh}}, \bibinfo {author} {\bibfnamefont {V.~B.}\ \bibnamefont {Verma}}, \bibinfo {author} {\bibfnamefont {S.~W.}\ \bibnamefont {Nam}}, \bibinfo {author} {\bibfnamefont {F.}~\bibnamefont {Marsili}}, \bibinfo {author} {\bibfnamefont {A.}~\bibnamefont {Ferrier}}, \bibinfo {author} {\bibfnamefont {P.}~\bibnamefont {Goldner}}, \bibinfo {author} {\bibfnamefont {H.}~\bibnamefont {Herrmann}}, \bibinfo {author} {\bibfnamefont {C.}~\bibnamefont {Silberhorn}}, \bibinfo {author} {\bibfnamefont {W.}~\bibnamefont {Sohler}}, \bibinfo {author} {\bibfnamefont {M.}~\bibnamefont {Afzelius}}, \ and\ \bibinfo {author} {\bibfnamefont {N.}~\bibnamefont {Gisin}},\ }\href {\doibase 10.1038/nphoton.2014.215} {\bibfield  {journal} {\bibinfo  {journal} {Nature
  Photonics}\ }\textbf {\bibinfo {volume} {8}},\ \bibinfo {pages} {775} (\bibinfo {year} {2014})}\BibitemShut {NoStop}%
\bibitem [{\citenamefont {Hucul}\ \emph {et~al.}(2015)\citenamefont {Hucul}, \citenamefont {Inlek}, \citenamefont {Vittorini}, \citenamefont {Crocker}, \citenamefont {Debnath}, \citenamefont {Clark},\ and\ \citenamefont {Monroe}}]{Hucul2015}%
  \BibitemOpen
  \bibfield  {author} {\bibinfo {author} {\bibfnamefont {D.}~\bibnamefont {Hucul}}, \bibinfo {author} {\bibfnamefont {I.~V.}\ \bibnamefont {Inlek}}, \bibinfo {author} {\bibfnamefont {G.}~\bibnamefont {Vittorini}}, \bibinfo {author} {\bibfnamefont {C.}~\bibnamefont {Crocker}}, \bibinfo {author} {\bibfnamefont {S.}~\bibnamefont {Debnath}}, \bibinfo {author} {\bibfnamefont {S.~M.}\ \bibnamefont {Clark}}, \ and\ \bibinfo {author} {\bibfnamefont {C.}~\bibnamefont {Monroe}},\ }\href@noop {} {\bibfield  {journal} {\bibinfo  {journal} {Nature Physics}\ }\textbf {\bibinfo {volume} {11}},\ \bibinfo {pages} {37} (\bibinfo {year} {2015})}\BibitemShut {NoStop}%
\bibitem [{\citenamefont {Monz}(2011)}]{monz2011quantum}%
  \BibitemOpen
  \bibfield  {author} {\bibinfo {author} {\bibfnamefont {T.}~\bibnamefont {Monz}},\ }\href@noop {} {Ph.D. thesis},\ \bibinfo  {school} {University of Innsbruck} (\bibinfo {year} {2011})\BibitemShut {NoStop}%
\bibitem [{\citenamefont {Janacek}(2015)}]{Hugothesis}%
  \BibitemOpen
  \bibfield  {author} {\bibinfo {author} {\bibfnamefont {H.}~\bibnamefont {Janacek}},\ }\href@noop {} {Ph.D. thesis},\ \bibinfo  {school} {University of Oxford} (\bibinfo {year} {2015})\BibitemShut {NoStop}%
\bibitem [{\citenamefont {Jung}\ \emph {et~al.}(2009)\citenamefont {Jung}, \citenamefont {Brambilla},\ and\ \citenamefont {Richardson}}]{jung2009comparative}%
  \BibitemOpen
  \bibfield  {author} {\bibinfo {author} {\bibfnamefont {Y.}~\bibnamefont {Jung}}, \bibinfo {author} {\bibfnamefont {G.}~\bibnamefont {Brambilla}}, \ and\ \bibinfo {author} {\bibfnamefont {D.~J.}\ \bibnamefont {Richardson}},\ }\href@noop {} {\bibfield  {journal} {\bibinfo  {journal} {Optics express}\ }\textbf {\bibinfo {volume} {17}},\ \bibinfo {pages} {16619} (\bibinfo {year} {2009})}\BibitemShut {NoStop}%
\bibitem [{\citenamefont {Zhang}\ \emph {et~al.}(2025)\citenamefont {Zhang}, \citenamefont {Phillips}, \citenamefont {Monga}, \citenamefont {Saglamyurek}, \citenamefont {Wu},\ and\ \citenamefont {Haeffner}}]{berkeley_freq_qubits}%
  \BibitemOpen
  \bibfield  {author} {\bibinfo {author} {\bibfnamefont {C.}~\bibnamefont {Zhang}}, \bibinfo {author} {\bibfnamefont {J.}~\bibnamefont {Phillips}}, \bibinfo {author} {\bibfnamefont {I.}~\bibnamefont {Monga}}, \bibinfo {author} {\bibfnamefont {E.}~\bibnamefont {Saglamyurek}}, \bibinfo {author} {\bibfnamefont {Q.}~\bibnamefont {Wu}}, \ and\ \bibinfo {author} {\bibfnamefont {H.}~\bibnamefont {Haeffner}},\ }\href {https://arxiv.org/abs/2503.05014} {} (\bibinfo {year} {2025}),\ \Eprint {http://arxiv.org/abs/2503.05014} {arXiv:2503.05014 [quant-ph]} \BibitemShut {NoStop}%
\bibitem [{\citenamefont {Connell}\ \emph {et~al.}(2021)\citenamefont {Connell}, \citenamefont {Scarabel}, \citenamefont {Bridge}, \citenamefont {Shimizu}, \citenamefont {Blūms}, \citenamefont {Ghadimi}, \citenamefont {Lobino},\ and\ \citenamefont {Streed}}]{streed_freq_qubits}%
  \BibitemOpen
  \bibfield  {author} {\bibinfo {author} {\bibfnamefont {S.~C.}\ \bibnamefont {Connell}}, \bibinfo {author} {\bibfnamefont {J.}~\bibnamefont {Scarabel}}, \bibinfo {author} {\bibfnamefont {E.~M.}\ \bibnamefont {Bridge}}, \bibinfo {author} {\bibfnamefont {K.}~\bibnamefont {Shimizu}}, \bibinfo {author} {\bibfnamefont {V.}~\bibnamefont {Blūms}}, \bibinfo {author} {\bibfnamefont {M.}~\bibnamefont {Ghadimi}}, \bibinfo {author} {\bibfnamefont {M.}~\bibnamefont {Lobino}}, \ and\ \bibinfo {author} {\bibfnamefont {E.~W.}\ \bibnamefont {Streed}},\ }\href {\doibase 10.1088/1361-6455/ac2984} {\bibfield  {journal} {\bibinfo  {journal} {Journal of Physics B: Atomic, Molecular and Optical Physics}\ }\textbf {\bibinfo {volume} {54}},\ \bibinfo {pages} {175503} (\bibinfo {year} {2021})}\BibitemShut {NoStop}%
\bibitem [{\citenamefont {Gao}\ \emph {et~al.}(2023)\citenamefont {Gao}, \citenamefont {Blackmore}, \citenamefont {Hughes}, \citenamefont {Doherty},\ and\ \citenamefont {Goodwin}}]{goodwin_cavity_proposal}%
  \BibitemOpen
  \bibfield  {author} {\bibinfo {author} {\bibfnamefont {S.}~\bibnamefont {Gao}}, \bibinfo {author} {\bibfnamefont {J.~A.}\ \bibnamefont {Blackmore}}, \bibinfo {author} {\bibfnamefont {W.~J.}\ \bibnamefont {Hughes}}, \bibinfo {author} {\bibfnamefont {T.~H.}\ \bibnamefont {Doherty}}, \ and\ \bibinfo {author} {\bibfnamefont {J.~F.}\ \bibnamefont {Goodwin}},\ }\href {\doibase 10.1103/PhysRevApplied.19.014033} {\bibfield  {journal} {\bibinfo  {journal} {Phys. Rev. Appl.}\ }\textbf {\bibinfo {volume} {19}},\ \bibinfo {pages} {014033} (\bibinfo {year} {2023})}\BibitemShut {NoStop}%
\bibitem [{\citenamefont {Magesan}(2011)}]{fidelity_channel_param}%
  \BibitemOpen
  \bibfield  {author} {\bibinfo {author} {\bibfnamefont {E.}~\bibnamefont {Magesan}},\ }\href@noop {} {\bibfield  {journal} {\bibinfo  {journal} {Quantum Info. Comput.}\ }\textbf {\bibinfo {volume} {11}} (\bibinfo {year} {2011})}\BibitemShut {NoStop}%
\end{thebibliography}%

\end{document}